\RequirePackage{rotating}
\pdfoutput=1
\documentclass[fleqn,usenatbib,usedcolumn]{mnras}

\usepackage{savesym}
\usepackage{amsmath}
\savesymbol{iint}
\usepackage{txfonts}
\restoresymbol{TXF}{iint}

\volume{{\rm in press}}

\usepackage[T1]{fontenc}
\usepackage{ae,aecompl}

\usepackage{graphicx}	
\usepackage{amssymb}	
\usepackage{subfig}
\usepackage{rotating}
\usepackage{float}

\AtBeginDocument{%
 \abovedisplayskip=3pt plus 3pt minus 9pt
 \abovedisplayshortskip=0pt plus 3pt
 \belowdisplayskip=12pt plus 3pt minus 9pt
 \belowdisplayshortskip=7pt plus 3pt minus 4pt
}

\newcommand\Tstrut{\rule{0pt}{3.6ex}}         
\newcommand\Bstrut{\rule[-1.9ex]{0pt}{0pt}}   

\title[Dynamics of Cluster Starbursts]{The Spatially Resolved Dynamics
  of Dusty Starburst Galaxies in a $z$\,$\sim$\,0.4 Cluster: Beginning the Transition from Spirals to S0s}

\author[H. L. Johnson et al.]{\parbox[h]{\textwidth}{
H.\,L.\ Johnson$^{1}$\thanks{E-mail: h.l.johnson@dunelm.org.uk},
C.\,M.\ Harrison$^{1}$,
A.\,M.\ Swinbank$^{2,1}$,
R.\, G.\ Bower$^{2,1}$,
Ian Smail$^{1,2}$,
Y.\ Koyama$^{3}$,
J.\,E.\ Geach$^{4}$}
\vspace*{4pt}\\
$^{1}$Center for Extragalactic Astronomy, Department of Physics, Durham University, South Road, Durham, DH1 3LE, UK\\
$^{2}$Institute for Computational Cosmology, Department of Physics, Durham University, South Road, Durham, DH1 3LE, UK \\
$^{3}$Subaru Telescope, National Astronomical Observatory of Japan, 650 North A'ohoku Place, Hilo, HI 96720, USA\\
$^{4}$Centre for Astrophysics Research, Science and Technology Research Institute, University of Hertfordshire, Hatfield, AL10 9AB, UK \\
\vspace{-0.6cm}
}

\date{Accepted XXX. Received YYY; in original form ZZZ}

\pubyear{2016}

\begin{document}
\label{firstpage}
\pagerange{\pageref{firstpage}--\pageref{lastpage}}
\maketitle

\begin{abstract}
To investigate what drives the reversal of the
morphology--density relation at intermediate/high redshift,
we present a multi-wavelength analysis of 27 dusty
starburst galaxies in the massive cluster Cl\,0024+17 at $z$\,=\,0.4.
We combine H$\alpha$ dynamical maps from the VLT\,/\,FLAMES multi-IFU system with far-infrared imaging using \emph{Herschel}\,/\,SPIRE
and millimeter spectroscopy from IRAM\,/\,NOEMA, in order to measure the
dynamics, star formation rates and gas masses of this sample.
Most galaxies appear to be rotationally supported, with a
median ratio of rotational support to line-of-sight velocity
dispersion $v/\sigma\sim$\,5\,$\pm$\,2, and specific angular momentum
$\lambda_{\rm R}$\,=\,0.83\,$\pm$\,0.06 -- comparable to field
spirals of a similar mass at this redshift.  The star formation rates of
3--26\,M$_{\odot}$\,yr$^{-1}$ and average $^{12}$CO-derived gas mass of
$\sim$\,1\,$\times$\,10$^{10}$\,M$_{\odot}$ suggest gas depletion
timescales of $\sim$\,1\,Gyr ($\sim$\,0.25 of the cluster
crossing time). We derive characteristic dust
temperatures (mean $T_{\rm d}$\,=\,26\,$\pm$\,1\,K) consistent with
local galaxies of similar far-infrared luminosity, suggesting that the
low density gas is yet to be stripped.
Taken together, these results suggest that these starbursts have
only recently accreted from the field, with star
formation rates likely enhanced due to the effects of ram pressure.
In order to make the transition to cluster S0s these galaxies must lose $\sim$\,40\% of their specific angular momentum.  We suggest this must occur $\geq$\,1\,Gyr later, after the molecular gas has been depleted and/or stripped, via multiple tidal interactions with other cluster members.
\end{abstract}




\section{Introduction}

Rich clusters present a unique laboratory for studying the interaction between galaxies and their local environment. It has long been established that strong trends exist between the morphology, gas content and star formation of cluster galaxies, and the density of the neighbourhood in which they reside. The populations of rich clusters at $z$\,=\,0 are dominated by passive, gas-poor ellipticals and S0s, with star formation all but extinguished in central regions (the morphology density relation e.g.\,\citealt{dressler1980, bower1992, lewis2002, kodama2004, bamford2009}). However, observations of clusters at intermediate redshift show a striking increase in the fraction of blue star-forming galaxies in cluster cores, from almost zero in the present day to $\sim$\,20\% by z\,$\sim$\,0.4 \citep[e.g.\ ][]{butcheroemler1978}. This evolution is accompanied by another key evolutionary change: a sharp decline in the proportion of S0 galaxies (\citealt{dressler1997, poggianti2009}). Studies of emission line and luminous infrared-selected galaxies out to $z\sim$\,1.5 have confirmed that star-forming galaxies in fact make up the majority of the population in high redshift clusters \citep{tran2010, smail2014}. This increase in the number of star-forming galaxies and decrease of cluster S0s with increasing redshift strongly implies that the two populations are linked in an evolutionary scenario. 

An important realisation was that a large fraction of quiescent cluster members have suffered starburst activity in the recent past. Spectroscopic surveys have identified a significant population of k+a or "post-starburst" galaxies which appear to be more prevalent with increasing redshift (e.g. \citealt{couch1987, poggianti1999, pracy2005, tran2007, swinbank2007, delucia2009, pino2014}). Their spectra show strong Balmer absorption lines associated with the recent formation of massive A stars ($\leq$\,1Gyr ago), but a lack of emission lines suggesting the burst was rapidly quenched (e.g. \citealt{poggianti1999, poggianti2000}). However, difficulties arise when attempting to link these galaxies to the formation of S0s. Insufficient numbers of strong starbursts are detected in the optical to explain the post-starburst phase, and the luminosities of S0s are in fact substantially brighter than the proposed progenitor spirals \citep{poggianti1999,kodama2004,burstein2005,sandage2005}. These problems could both be solved if a considerable fraction of starburst activity is heavily obscured. Indeed, deep mid-infrared observations of intermediate redshift clusters with {\it Spitzer} and
{\it Herschel} have revealed an abundance of such dusty star-forming
galaxies which are missing from optical studies (e.g.\,\citealt{coia2005, geach2006, geach2009, elbaz2007,
  marcillac2007, koyama2008, oemler2009, kocevski2011, alberts2014}). In an era in which clusters were still accreting much of their mass, there not only appears to be a significant population of star-forming galaxies, but also many with their star formation temporarily enhanced in this dense environment.
  
In a model where spirals transition to S0s, the star formation of infalling galaxies must be rapidly quenched, their gas disks stripped, and the dynamics transform from rotationally supported disks (high angular momentum) to dispersion dominated spheroids (low angular momentum). Several authors also suggest that the bulge luminosity luminosity and bulge to disk ratio of S0s is too large for them to have evolved from spirals by disk fading alone (\citealt{dressler1980, simien1986,kodama_smail2001,christlein_zabludoff2004,cortesi2013}), and spectral decomposition of local lenticulars has revealed that bulges have younger and more metal-rich stellar populations than their adjacent disk \citep{johnston2014}. It may be that a final, circumnuclear starburst is a critical stage in the transition between infalling spirals and passive S0s. The observational challenge is to identify the processes which may drive this transformation.

Several potential mechanisms are usually invoked to explain galaxy transformations in local clusters: interactions with the intra-cluster medium (ICM) such as ram pressure stripping, strangulation and thermal evaporation (\citealt{kenney2004, mccarthy2008, merluzzi2013, fumagalli2014, peng2015}), and tidal interactions, harassment, minor mergers or halo stripping (\citealt{mastropietro2005, bekki2009, smith2010, eliche2012, bialas2015}). What remains elusive is the relative contribution of each process, and an understanding of how the timescales they operate on vary with cluster mass and size (hence redshift), as well as the stellar, halo and gas mass of the infalling galaxy \citep[e.g.\ ][]{boselligavazzi2006, boselligavazzi2014}. 

It is important to study cluster galaxies over a range in redshift to thoroughly explore the mechanisms described above, and to understand what drives the reversal of the morphology--density relation. For example, due to the increased gas fractions of galaxies at high redshift (e.g. \citealt{tacconi2010,geach2011}), the initial compression of the
ISM may be more likely to enhance the star formation \citep{quilis2000,hopkins2006,bekki2014,sales2015}. Due to the lower mass of typical clusters (compared to those at $z\sim$\,0), the lower ram pressure from the intra-cluster medium may also result in the starbursts being both more intense, and long lived. Tidal forces from increased galaxy-galaxy interaction rates in rapidly assembling clusters may also destablise the gas disks, causing a burst of star formation, and a morphological and dynamical transformation.

Identifying cluster starbursts and measuring their dynamics, star-formation rates and molecular gas properties appears to be key to unravelling the complexities of galaxy evolution in clusters. The short lifetime of the starburst activity provides a snapshot of galaxies which may be undergoing a transition, allowing us to search for potential triggers. For example, asymmetric gas disks may provide evidence for ram pressure
stripping \citep{bekki2014}, high dust temperatures (compared to
galaxies in the field) may imply that the cold gas/dust has
been stripped (e.g. \citealt{rawle2012}), whilst galaxy-galaxy interactions (mergers) may result in complex kinematic signatures \citep{mihos1998,colina2005} depending on the interaction stage and nature of the system \citep{bellochi2013,hung2016}. Complementing the dynamics with observations of molecular gas also
allows us to infer the likely timescales for the starbursts and
measure how long the starburst can be maintained.

In this paper we present a multi-wavelength study of 28
spectroscopically confirmed, 24\,$\mu$m-bright galaxies within
Cl\,0024+17. This pilot study forms part of a wider investigation into the properties of cluster starburst galaxies. Analysis of our full sample of $\sim$\,150 galaxies observed with FLAMES\,/\,KMOS, from ten massive clusters between
$0.2<z<1.5$, will be presented in paper {\sc ii}. We measure the
morphologies and dynamics of the galaxies using \emph{HST} and
VLT\,/\,FLAMES multi-IFU observations respectively,
infer star-formation rates from far-infrared observations with
\emph{Herschel}\,/\,SPIRE and measure molecular gas masses using
$^{12}$CO(1--0) emission from NOEMA. As an original ``Butcher \&
Oemler'' cluster, Cl\,0024+17 has a significant population of
blue, star-forming galaxies, and is one of the best studied clusters
at intermediate redshift ($z$\,=\,0.395), with a virial mass $M_{\rm
  vir}$\,=\,(1.2\,$\pm$\,0.2)\,$\times$\,10$^{15}$\,M$_{\odot}h^{-1}$
\citep{umetsu2010}, and X-ray luminosity of $L_{\rm
  X}$\,$\sim$\,2.9\,$\times$\,10$^{44}$\,erg\,s$^{-1}$
\citep{zhang2005}.  With a rich variety of star forming galaxies in
the cluster, and multi-wavelength ancillary data, Cl\,0024 provides a
useful pilot study for investigating the properties of dusty
starbursts in galaxy clusters. 

The structure of this paper is as follows. In \S2 we describe the
target selection, observations and data reduction.  In \S3 we describe
the galaxy integrated properties, such as star formation rates,
stellar and gas masses, and dust temperatures.  In \S4 we describe the
internal proprties of the galaxies from the IFU observations:
H$\alpha$ dynamics, two-dimensional maps of star formation, rotation
velocity and line of sight velocity dispersion.  We then explore the
properties of these dusty starbursts in the context of their
environment, comparing to field spirals and local lenticulars and
searching for trends as a function of cluster radius.  We consider
which mechanisms may have already acted upon these galaxies, and
discuss what remains to be achieved to complete the transition to S0s,
in \S5. Finally \S6 summarises our main results.  Throughout this
paper we assume a $\Lambda$CDM cosmology with parameters
$\Omega_m$\,=\,0.27, $\Omega_{\Lambda}$\,=\,0.73 and
$H_{0}$\,=\,72\,km\,s$^{-1}$\,Mpc$^{-1}$. The average seeing for our
IFU observations, 0.5$''$, corresponds to a physical scale of
2.6\,kpc at $z$\,$\sim$\,0.4.  Unless otherwise stated, all magnitudes are quoted on the
AB system.

%
%
\section{Target Selection, Observations \& Data Reduction}
\subsection{Cl\,0024+17}

Cl\,0024+17 is a well studied cluster with extensive archival
multi-wavelength imaging and spectroscopy, and its large population of blue star-forming galaxies makes it an ideal environment for exploring the properties of cluster starbursts. In
optical wavelengths Cl\,0024 appears fairly unremarkable, with a well
concentrated mass profile and relatively little sub-structure
(Fig. \ref{fig:spatial_dist}), but it is thought this disguises a rather eventful dynamical history. Spectroscopic observations reveal two distinct
components in the line of sight: a dominant ``cluster'' which has a
velocity dispersion of $\sim$\,1000\,km\,s$^{-1}$, and a
foreground sub-group offset by
$\Delta v\sim$\,3000\,km\,s$^{-1}$ which has a velocity dispersion of
$\sim$\,500\,km\,s$^{-1}$ (Fig.~\ref{fig:sample_properties}). The sub-group appears to have undergone a
high speed collision with the cluster
around $\sim$\,3\,Gyr ago \citep{czoske2001,czoske2002}. 

\citet{czoske2002} explored the proposed cluster - sub-group collision via numerical simulations of
dark matter haloes, concluding that this scenario could explain a well documented
discrepancy between mass estimates derived from lensing
(e.g. \citealt{comerford2006,hoekstra2007,zitrin2009,umetsu2010}),
velocity dispersion \citep{diaferio2005} and X-ray studies
(\citealt{soucail2000,ota2004,zhang2005}). Cl\,0024 has a lower X-ray flux and lower central velocity dispersion than expected for a virialised system, and this leads to some interesting implications. Processes such as galaxy harassment are usually most effective in the outskirts of clusters, or within groups as a method of ``pre-processing'', since the lower relative velocity of galaxies leads to longer interaction times. However \citet{moran2007} suggest that for Cl\,0024, harassment may be effective down to surprisingly small radii, with ram pressure stripping weak until $\sim$\,300\,kpc from the core. Passive spirals appear to be relatively long-lived (1--2\,Gyr; \citealt{treu2003}) with galaxies experiencing a slower transition due to tidal interactions. In this respect Cl\,0024 makes a useful analog to higher redshift clusters which are still in the process of assembling.

\subsection{Target Selection}

To identify a sample of dust obscured cluster starbursts suitable for
VLT\,/\,FLAMES IFU observations, we exploit the
\emph{Spitzer}\,/\,MIPS 24\,$\mu$m imaging from \citet{geach2006} and
select bright mid-infrared bright sources with 24$\mu$m fluxes of
$S_{\rm 24\mu m}$\,=\,0.15--1\,mJy (median $S_{\rm 24\mu
  m}$\,=\,0.33\,$\pm$\,0.03\,mJy) which have also been
spectroscopically confirmed as cluster members \citep{moran2005}.
Approximately two-thirds of the sample are known H$\alpha$ or [O{\sc
    ii}] emitters, as identified from Suprime-Cam narrowband H$\alpha$
imaging and DEIMOS spectroscopy \citep{kodama2004,moran2005}. We also include two H$\alpha$ emitters of unknown 24 $\mu$m flux (IDs 9 and 10) which we used to fill vacant IFUs. We find no further distinction between these two galaxies and the rest of the sample, suggesting they are cluster starbursts which are simply not as dust obscured.

\begin{figure}
\includegraphics[width=0.48\textwidth]{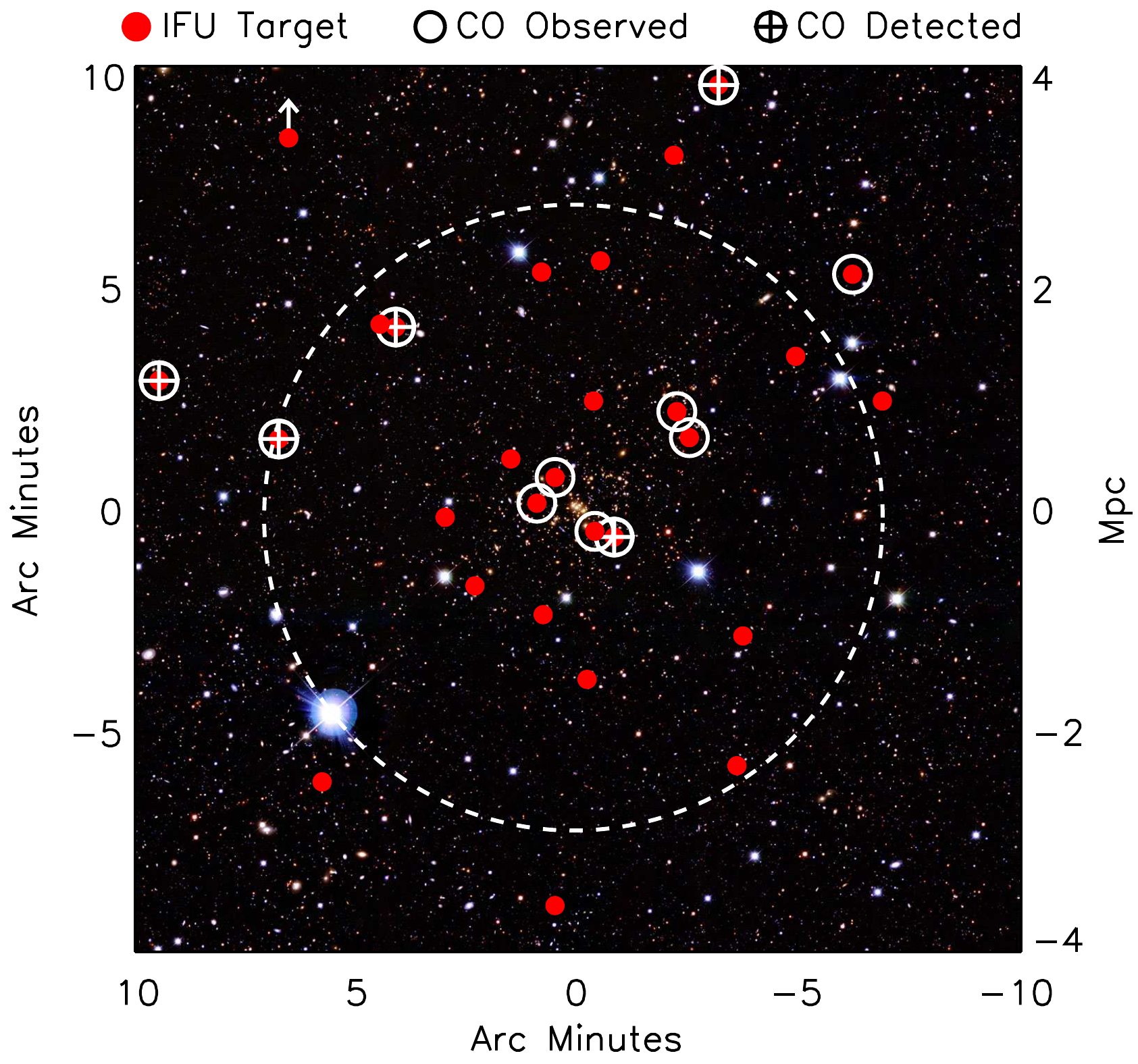}
\caption{SUBARU SuprimeCam $BVI$-colour image of Cl\,0024+17, centred on 
  $\alpha$:\,00:26:36.0 $\delta$:\,+17:08:36 (J2000).  The dashed line
  represents the virial radius of the cluster, R$_{\rm
    vir}$\,=\,1.7\,Mpc.  We highlight the 28 spectroscopically
  confirmed dusty starbursts which were observed using
  the FLAMES multi object IFU. Our targets have projected cluster-centric radii of
  0.2\,--\,3.9\,Mpc.  We circle galaxies also observed using IRAM
  PdBI, with crosses to indicate $>$5\,$\sigma$ $^{12}$CO(1--0)
  detections. It appears that galaxies further from the cluster centre are perhaps more likely to be detected in CO (also see Fig.~\ref{fig:sample_properties}).}
\label{fig:spatial_dist}
\end{figure}

\begin{figure*}%
  \centering
  \subfloat{{\includegraphics[width=0.44\textwidth]{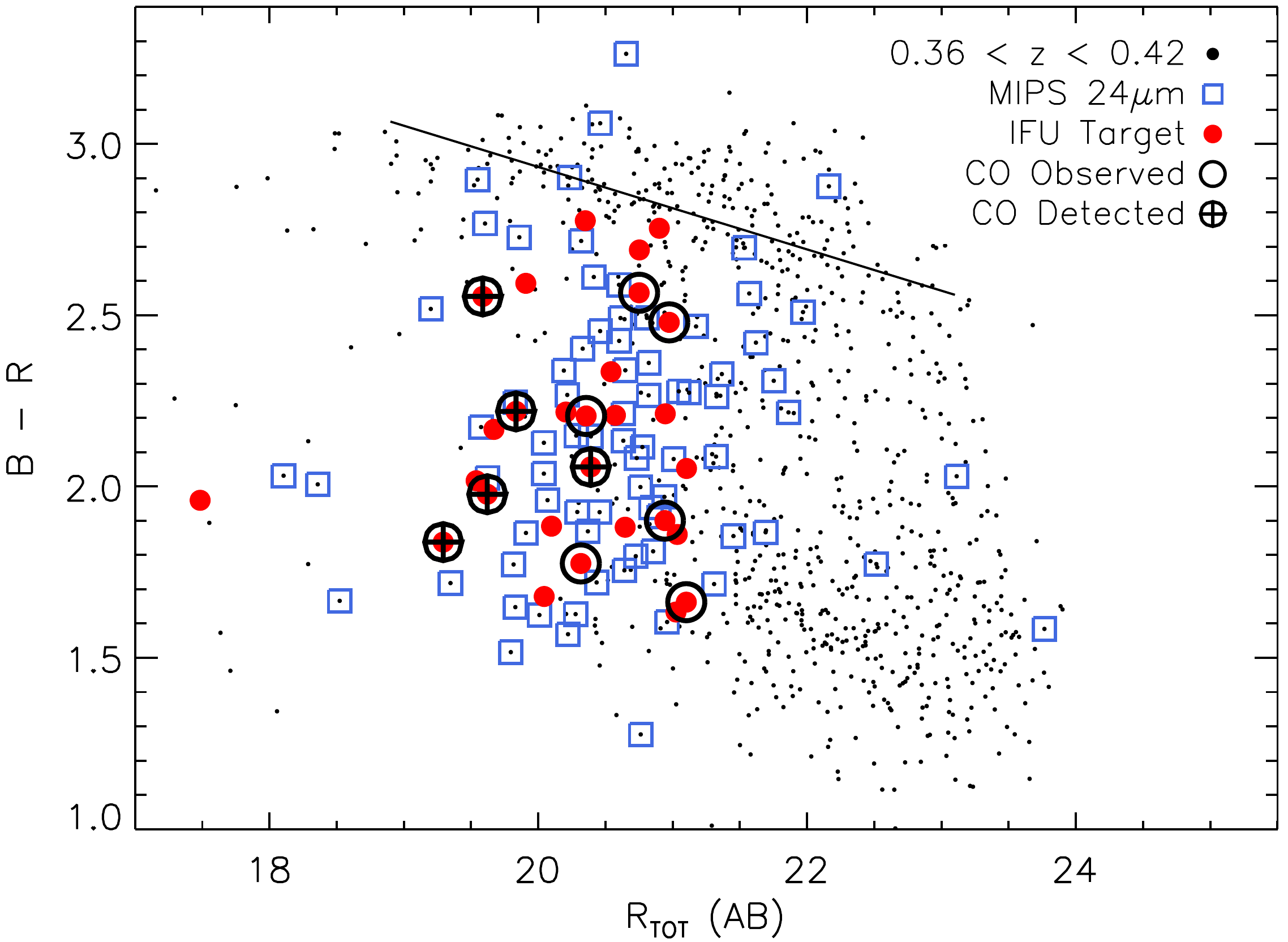} }}%
  \qquad
  \subfloat{{\includegraphics[width=0.47\textwidth]{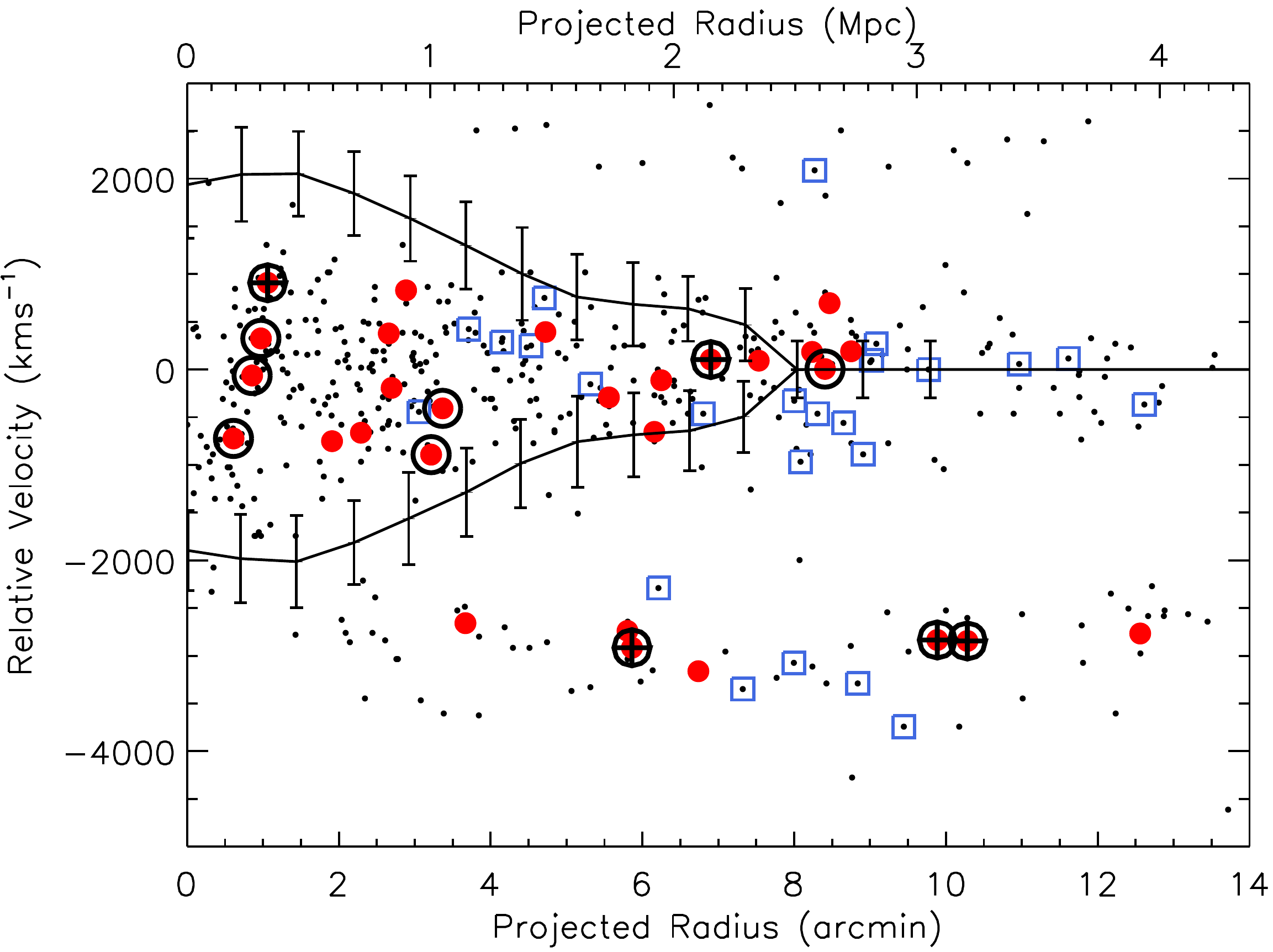} }}
  \caption{Properties of our starburst sample. {\it Left:}
    Colour-magnitude relation for galaxies within a 9\,arcmin radius
    of the Cl\,0024+17 cluster centre and within $\Delta z$\,$<$\,0.03
    of the cluster redshift.  We identify the cluster red sequence
    (solid line). Many targets in our
    sample lie between the red sequence and blue cloud -- a
    consequence of dust-obscured star formation in blue, star-forming
    galaxies. {\it Right:} Line of sight velocity with respect to the
    cluster centre, versus projected cluster-centric radius.  Solid
    lines show the caustics which illustrate the escape velocity of
    the cluster \citep{diaferio2005}.  The structure at
    $\sim$\,$-$3000\,km\,s$^{-1}$ appears to be a group in the line of
    sight, which previously passed through Cl\,0024+17 \citep{czoske2002}.  Seven of our cluster starbursts lie within this foreground structure.}
  \label{fig:sample_properties}%
\end{figure*}

As shown by Fig.~\ref{fig:spatial_dist} \& \ref{fig:sample_properties}, our final starburst sample of 28 galaxies is well
dispersed throughout the cluster, with projected cluster-centric radii
of $R$\,=\,0.2\,--\,3.9\,Mpc. 20 of 28 galaxies lie within the cluster virial radius of R$_{\rm vir}$\,=\,1.7\,Mpc \citep{treu2003}. Seven of our targets are associated with
the foreground group, which will allow us to assess the effect of an
accelerated environment on the galaxy properties.

In Fig.~\ref{fig:sample_properties} we show the ($B$--$R$) colour for
all galaxies within the redshift range 0.36\,$\leq$\,$z$\,$\leq$\,0.42
(equivalent to $\Delta v$\,$\sim$\,9000\,km\,s$^{-1}$).
No colour cut was made in selecting our IFU sample, and as can be
seen from this figure, the MIPS\,24$\mu$m detections and our IFU targets tend to lie either in the blue-cloud or between the blue-cloud and the
cluster red sequence (which can be identified
between ($B$--$R$)\,$\sim$\,2.5\,--\,3.2). Some starbursts likely show redder colours due to the influence of dust
on a population of intrinsically blue and star-forming galaxies.

In our analysis we also exploit archival observations of Cl\,0024+17
which were taken with {\it HST}\,/\,WFPC2 as a 39-point sparse mosaic \citep{treu2003}.  The images
were taken with the F814W filter (corresponding to rest-frame 580\,nm)
and cover approximately one-third of our sample (Fig. \ref{fig:dynamics_summary1}). We find a high incidence of disks with a range of bulge to disk ratios, and none of the galaxies appear to be undergoing a major merger. We note that there is no obvious trend in morphology with projected cluster-centric radius or association with the cluster or foreground group.

\subsection{FLAMES IFU Observations}
Observations of our 28 starbursts were made using
the FLAMES multi-object IFU system on the VLT between December 13 2012
and November 29 2013 as part of ESO runs 089.A-0983 and 092.A-0135.
FLAMES employs 15 IFUs across a patrol field of 25 arcmin diameter
(Fig. \ref{fig:spatial_dist}) and each of the deployable integral
field units consists of a near rectangular array of 20 microlenses
(with pixel scale 0.52$''$\,$\times$\,0.52$''$), resulting in a total aperture
of $\sim$\,3$''$\,$\times$\,2$''$.  All of the observations were taken
in dark time and excellent seeing ($<\,$0.5$''$).  We used the GIRAFFE
spectrograph with the ${\rm LR}\,881.7$ filter to cover the H$\alpha$
emission line in all of our targets, which at $z\sim $\,0.39 is
redshifted to $\sim$\,9100\AA.  At this wavelength the spectral
resolution is $R$\,=\,$\lambda$\,/\,$\Delta\lambda$\,$\sim$\,9000 (as
measured from skylines), and we correct for this instrumental dispersion in all of our observations. Each observing block was split in to two
1.2\,ks observations, with sub-pixel dithers to improve the effective spatial sampling. The total on-source integration time per target was 3\,--\,4\,hrs.

To reduce the data we used the standard {\sc EsoRex} pipeline, which
extracts the fibres from each IFU, flatfields and wavelength
calibrates to form a set of preliminary datacubes.  We reduced each
observation individually, before sky subtracting and flux calibrating
the spectra.  For sky spectra we masked emission lines and continuum
from the IFUs and then calculated an average sky spectrum which was then removed from all targets.

To create a mosaic for each galaxy we apply the sub-pixel offsets from our dither patterns, which were verified
using the centroided continuum emission from the three brightest
galaxies in the sample. This resulted in an effective pixel scale of 0.17$''$. In combining the data cubes we used a 3$\sigma$
clipped average. The final spectra, integrated over each IFU, are shown in Appendix A along with
the {\it HST} (where available; \citealt{treu2003}) or Subaru imaging \citep{kodama2004}.  All but one of the targets were detected (and
spatially resolved), leaving us with a final sample of 27 cluster
starbursts. We note that the undetected target was the galaxy with the weakest H$\alpha$ emission in the parent catalog. We exclude this source from the rest of the analysis.

To create the two-dimensional H$\alpha$ emission, velocity, and
velocity dispersion maps, we first measure the systemic redshift by
fitting the H$\alpha$ and [N{\sc ii}]$\lambda\lambda$6548, 6583
emission lines in the collapsed spectra. We
then repeat this procedure on a pixel-by-pixel basis using a
$\chi^{2}$ minimisation procedure, inverse weighting the fit using a sky spectrum to account for increased noise at the positions of 
OH sky lines.  We fit the H$\alpha$ and [N{\sc ii}] doublet emission
lines simultaneously, allowing the centroid, intensity and width of the
Gaussian profile to vary. The FWHM of the H$\alpha$ and
[N{\sc ii}] lines are coupled and the intensity ratio of the [N{\sc
    ii}] $\lambda$6548\,/\,$\lambda$6583 was fixed at 2.98. We require a SNR $>$\,5 in order to
record a detection in each individual pixel. During
the fitting, we convolve the line profile with the instrumental
dispersion.  As such, all measurements are corrected for the
instrumental resolution.

Fig.~\ref{fig:poster_plot} shows the H$\alpha$ velocity field of each of our sample, and in
Fig.~\ref{fig:dynamics_summary1} we show the resolved H$\alpha$ emission, dynamics and line-of-sight velocity dispersion maps of all galaxies. Although we observe a range of H$\alpha$ morphologies, the majority of the sample appear to have regular (disk-like)
velocity fields with velocity dispersions which peak towards the dynamical centre. It is useful to note that in the galaxy integrated one-dimensional spectra (Fig. \ref{fig:onedspectra}), more than half show
emission lines profiles which are double-peaked. This is also indicative of disk-like dynamics, with either increased dust obscuration towards central regions, or ring-like emission. We will return to a more
detailed discussion of the dynamics in \S\ref{dynamics}.

\subsection{Plateau de Bure Observations}
\label{pdbi}
To assess the evolutionary state of our target galaxies we have also sought to obtain cold gas masses for a subset. This will allow us to compare the gas properties to field galaxies of a similar mass at the same redshift. We used the IRAM PdBI and its NOEMA upgrade to target the $^{12}$CO\,J(1\,$\rightarrow$\,0) transition in 11 galaxies, five of which were previously presented in \citet{geach2009,geach2011}. These inital targets were predominantly in the outskirts of Cl\,0024+17, with cluster-centric radii of
1.8\,--\,3.3\,Mpc, and so to complement this data we selected a further six galaxies which lie closer to the (projected) cluster core (Fig.~\ref{fig:spatial_dist}). Observations took place in 2014 June as part of programme S14BT. Both sets of observations analysed here
used the compact ``D" configuration using six or seven antennae.  We
targeted the $^{12}$CO\,J(1\,$\rightarrow$\,0) 115.27\,GHz rotational
transition, which $z$\,=\,0.395 is redshifted into the 3\,mm band
with $\nu_{\rm obs}$\,=\,82.63\,GHz. The central frequency of the
3\,mm receiver was set to coincide with the CO(1-0) line at the
spectroscopic redshift. For further details on the setup of the first set of observations, see \citet{geach2011}.

For the most recent sample, since one frequency setup was sufficient to cover all targets we
required only one phase calibrator, switching between three pointings.  The
correlator was set up with 2.5\,MHz spacing (2\,$\times$\,64 channels,
320\,MHz bandwidth), so as to accommodate any potential offset from
the systemic redshift or a broad emission line profile. To increase
observing efficiency, we chose pointings where two cluster starbursts
had sufficiently small on-sky separations that they lay within a single
primary beam. Exposure times were 4.15\,hr per source pair.  The data
was calibrated, mapped and analysed using the software {\sc gildas}
\citep{guilloteau2000}.

Combining our six new observations with those
from \citet{geach2011}, we detect the CO J(1\,$\rightarrow$\,0)
transition in five of eleven targets. We require 3$\sigma$ for
a detection, with upper limits on $L^{\prime}_{\rm CO}$ for
non-detections based on the rms noise and median line width of the sample. One dimensional spectra are shown in Appendix Fig.~\ref{fig:onedspectra}. For the five detections we find line fluxes in the range $f_{\rm CO(1-0)}$\,=\,(255\,--\,788)\,mJy\,km\,s$^{-1}$ and an average line width of $\sigma_{\rm CO}$\,=\,140\,$\pm$\,25\,km\,s$^{-1}$. To estimate the average flux of the sample we stack the CO spectra from all 11 galaxies. We find $f_{\rm CO(1-0)}$\,=\,309\,$\pm$\,30\,mJy\,km\,s$^{-1}$, and derive the same flux independent of whether we take an average, a median, or a noise weighted sum. Despite only one new
detection, stacking the data allows us to place important constraints on the gas properties of these starbursts. We will return to this in \S\ref{integrated}.

\section{Analysis \& Discussion: Galaxy Integrated Properties}
\label{integrated}

Before discussing the dynamics, star formation and molecular gas
properties of our cluster starburst sample, we first derive their
stellar masses and star formation rates using the multi-wavelength
imaging available for the cluster.  This allows the properties
(current and final stellar mass, baryon fraction, dynamical state, gas
depletion timescale) to be set in context with the field and cluster population at the same redshift.

\subsection{Stellar Masses}
\label{masses}
We begin by deriving absolute $K$-band magnitudes and stellar masses
for the galaxies in our sample, using eight-band photometry from the
optical to mid-infrared. The $BVRI$-band fluxes were taken from
\citet{czoske2002,treu2003} (which comprises Canada-France-Hawaii Telescope CFHT12k imaging). The  $J$- and $K$-band photometry is given in
\citet{kneib2003,smith2005}, and is extracted from observations made
with the WIRC camera on the Palomar Hale 200$''$ telescope. In the mid-infrared, we perform aperture photometry using archival {\it Spitzer} IRAC imaging in the 3.6 and 4.5\,$\mu$m bands. In all cases, we use 2.5$''$ aperture magnitudes and apply the appropriate aperture corrections to total magnitudes.

To fit the spectral energy distributions (SEDs) of each galaxy and
hence infer star-formation histories and stellar masses, we employ the
{\sc hyperz} fitting code \citep{bolzonella2000}.  Model SEDs
are characterised by their star formation history and
parameterised by age, reddening and redshift.  Using spectral
templates derived from the \citet{bruzual&charlot2003} evolutionary
code we consider six star formation histories: a single burst (B),
constant star formation (Im), and exponential decays of timescales 3,
5, 15 and 30\,Gyr (Sa,b,c,d respectively).  We use the
\citet{calzetti2000} reddening law and, given the obscured nature of
these cluster starbursts we allow $A_V$\,=\,0.0\,--\,2.5 magnitudes in
steps of $A_{\rm V}$\,=\,0.1. We find an average extinction across the sample of $A_V$\,=\,1.2\,$\pm$\,0.4 magnitudes. In each case we fix the redshifts of the synthetic
templates to match the H$\alpha$ emission.

We next integrate the best-fit star formation history to calculate the
stellar mass of each galaxy, accounting for mass loss from remnants
using the {\sc starburst99} synthesis models \citep{leitherer2011}.
Given the dusty nature of these galaxies, we note that there are
degeneracies between age and reddening which results in considerable
uncertainty in the stellar mass-to-light ratios, an issue that is likely to be compounded by complex star formation histories. To overcome some of these difficulties, we derive
the rest-frame $K$-band mass-to-light ratio a galaxy-by-galaxy basis, and then reapply the average value to the entire sample. The average mass to light ratio is $\left<\Upsilon_{\star}\right>$\,=\,$M_{\odot}$/$L_{\odot}^K$\,=\,0.35, and using this technique
we find galaxies in our sample to have stellar masses in the range of
M$_\star$\,=\,(1--10)\,$\times$\,10$^{10}$\,M$_{\odot}$, with a median of
$M_{\star}$\,=\,(2.8\,$\pm$\,0.3)\,$\times$\,10$^{10}$\,M$_{\odot}$. We will use these values in the rest of our analysis. Absolute $K$-band magnitudes and stellar masses (using the median $\Upsilon_{\star}$) are
given in Table~\ref{table:derived} of the Appendix.

To compare the stellar masses to those of field galaxies at the same
redshift, we exploit the COSMOS\,/\,UltraVISTA survey from
\citet{muzzin2013}, who derive a characteristic mass of
M$_\star$\,=\,(5.8\,$\pm$\,0.5)\,$\times$\,10$^{10}$\,M$_{\odot}$ at $z\sim$\,0.4. This suggests that, on average, the galaxies in our cluster starburst sample have slightly lower stellar masses than the ``typical'' M$_\star$ field
galaxy at this redshift.

\begin{figure}
 \includegraphics[width=0.48\textwidth, angle=0]{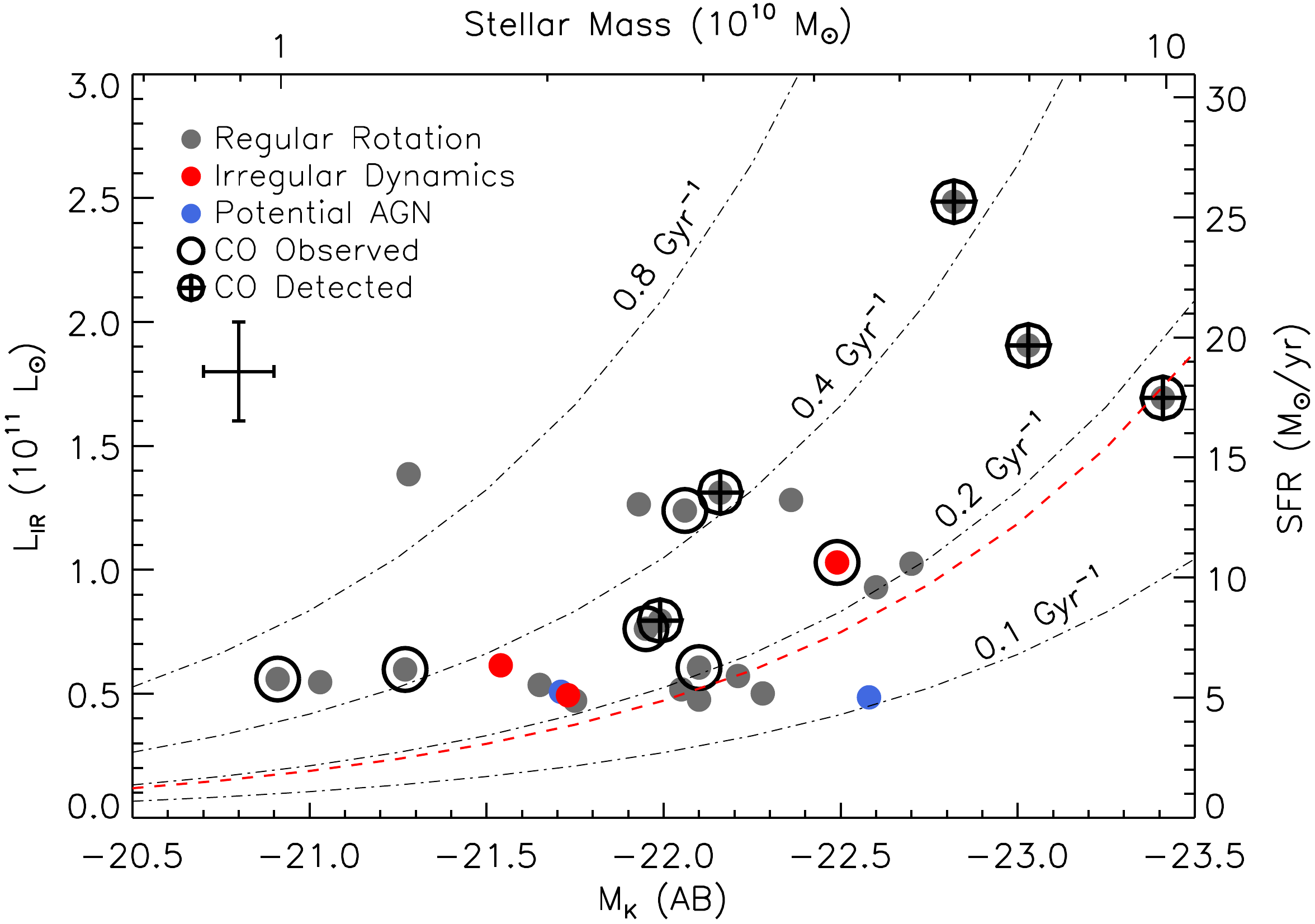}
  \caption{Infrared luminosity against $K$ band absolute magnitude,
    with additional axes to demonstrate how this relates to star
    formation rate and stellar mass. We use the median mass to
    light ratio for our sample of $M_{\odot}$/$L_{\odot}^{\rm K}$\,=\,0.35 (see \S\ref{masses}). Dashed lines indicate a constant
    specific star formation rate (sSFR), with a red line used to illustrate the sSFR of the so-called main-sequence at $z$\,=\,0.4. We highlight galaxies observed using
    PdBI / NOEMA, and find that detected galaxies tend to be of higher mass and
    brighter $L_{\rm IR}$. Red points are used to represent galaxies
    with dynamics which deviate from a rotating disk model, and blue points for galaxies which may have an AGN affect the dynamics (these classifications are assigned in \S\ref{dynamics}). We find no
    relationship between $L_{\rm IR}$, $M_{\rm K}$ and H$\alpha$
    dynamics.}
  \label{fig:group_sfr}
\end{figure}

\subsection{Star Formation Rates}
\label{sfr}

To estimate the star formation rates of these galaxies we adopt two
approaches, using the (reddening corrected) H$\alpha$ and the far-infrared imaging with
\emph{Herschel}. We detect strong H$\alpha$ emission in the
integrated spectra of all but one of our 28 IFU targets (which we have consequently dropped from the sample), with
luminosities in the range $L_{\rm
  H\alpha}$\,$\sim$\,$10^{40.6-41.8}$\,erg\,s$^{-1}$ (see
Fig. \ref{fig:dynamics_summary1} and Table \ref{table:observed}).  We
also make clear detections of the neighbouring [N{\sc
    ii}]$\lambda$6583 line, with ratios of $-$0.69\,$<$\,log([N{\sc
    ii}]/H$\alpha$)\,$<$\,0.04 and a median of $-$0.4\,$\pm$\,0.02. Since two of our sample exhibit log([N{\sc ii}]/H$\alpha$)\,$>$\,0, we discuss
the possibility of AGN contamination in \S\ref{dynamics}. We note that while the H$\alpha$ star formation rates of AGN may be unreliable, the far-infrared results are less likely to suffer contamination.

We use the integrated H$\alpha$ luminosity to
calculate star formation rate (SFR) estimates, applying the
calibration of \citet{kennicutt1998} for a Chabrier IMF (SFR\,M$_{\odot}$\,yr$^{-1}$\,=\,4.6\,$\times$\,10$^{-42}$\,L$_{\rm H\alpha}$\,erg\,s$^{-1}$). Using the attenuation, $A_V$, returned by {\sc
  hyperz} we also apply the dual reddenning law of
\citet{wuyts2013} which assumes
that the nebular emission is more attenuated than the stellar
continuum (since young, ionising stars will typically reside in
dustier regions). Therefore applying dust corrections of an average $A_{\rm gas}$\,=\,1.8\,$\pm$\,0.5 magnitudes, we estimate H$\alpha$ star formation rates of SFR$_{\rm H\alpha}$\,=\,0.3\,--\,16\,M$_{\odot}$\,yr$^{-1}$.

To compare far-infrared star formation rates to those derived from H$\alpha$ emission, we next exploit the {\it Herschel} PACS\,/\,SPIRE imaging
of Cl\,0024+17.  For PACS, deblended catalogs are available for the cluster core \citep{lutz2011}, which covers five of the galaxies in
our sample.  To derive 250, 350 and 500$\mu$m flux densities, we
use SPIRE imaging from the HerMES legacy programme \citep{oliver2012}.
We first align the astrometry of the images by stacking at the
24$\mu$m positions, applying $<$1$''$ shifts in $\Delta$RA and
$\Delta$Dec, and then deblend following the procedure of
\citet{swinbank2014}, using the {\it Spitzer} 24\,$\mu$m sources as
priors.  From our sample of 27 cluster starbursts, 20 are detected
above 11\,mJy at 250$\mu$m. The median 250, 350 and 500$\mu$m fluxes for the detections in each band are
20\,$\pm$\,3, 11\,$\pm$\,2 and 9\,$\pm$\,3\,mJy, respectively.

To derive far infrared luminosities we fit the far-infrared flux
densities at the spectroscopic redshift, using the far-infrared
templates from the \citet{chary2001}, \citet{draine2007} and
\citet{rieke2009} SED libraries.  At the cluster redshift PAH and
silicate features lie bluewards of the 24\,$\mu$m band, and so we
include all photometry between 24\,$\mu$m and 500\,$\mu$m.
Integrating the best fit SEDs we then derive bolometric luminosities of
$L_{\rm IR}$\,=\,(0.47\,--\,2.47)\,$\times$\,10$^{11}$\,L$_{\odot}$,
corresponding to star formation rates of SFR$_{\rm
  IR}$\,=\,3\,--\,26\,M$_{\odot}$yr$^{-1}$.   Far-infrared luminosities and star formation rates are given in
Table~\ref{table:derived}.

For galaxies with many upper limits on their far-infrared fluxes, we take the H$\alpha$ star formation rate and multiply this by the average SFR$_{\rm IR}$/SFR$_{\rm H\alpha}$ ratio, to estimate SFR$_{\rm IR}$. On average, the star formation rate derived from the far-infrared is a
factor 2.2\,$\pm$0.4 larger than that derived from the H$\alpha$, even
after applying a dust correction to the H$\alpha$ luminosity. This is equivalent to an additional A$_{\rm gas}$\,=\,0.9\,$\pm$\,0.2 (or A$_{\rm
  V}$\,=\,1.3\,$\pm$\,0.2). This offset may arise due to the
differing structures between the dust and stars, particularly if the
starbursts are centrally concentrated in an otherwise extended disk. Since the SED fitting provides a luminosity weighted result,
the attentuation predicted from {\sc hyperz} will be dominated by the extended disk and so systematically low. As a proxy for the
star formation rate, in the rest of our analysis we will use the
values derived from the far-infrared fluxes.

In Fig. \ref{fig:group_sfr} we plot infra-red luminosity versus
absolute $K$ band magnitude for the galaxies in our sample, converting
the values also to star formation rate and stellar mass, respectively.
We derive specific star formation rates (sSFR) in the range
(0.1\,--\,0.9)\,M$_{\odot}$Gyr$^{-1}$ with a median of
sSFR\,=\,0.27\,$\pm$\,0.03\,Gyr$^{-1}$. Our cluster
galaxies (excluding one AGN candidate, see \S4) lie on or above the so-called ``main sequence'' at
$z$\,$\sim$\,0.4 \citep{elbaz2007,noeske2007,karim2011}. On average
the galaxies are offset from the main-sequence by a factor of
1.8\,$\pm$\,0.2 at a fixed stellar mass, and hence we adopt the term ``starburst''. In \S\ref{dynamics} we assign each galaxy a dynamical classification -- rotationally supported or irregular -- based on their velocity map and line of sight dispersion map. In Fig.~\ref{fig:group_sfr} we split the sample by classifications and do not see
any strong trends between the star formation rate and the dynamical state of the galaxy.

\begin{figure}
 \includegraphics[width=0.45\textwidth, angle=0]{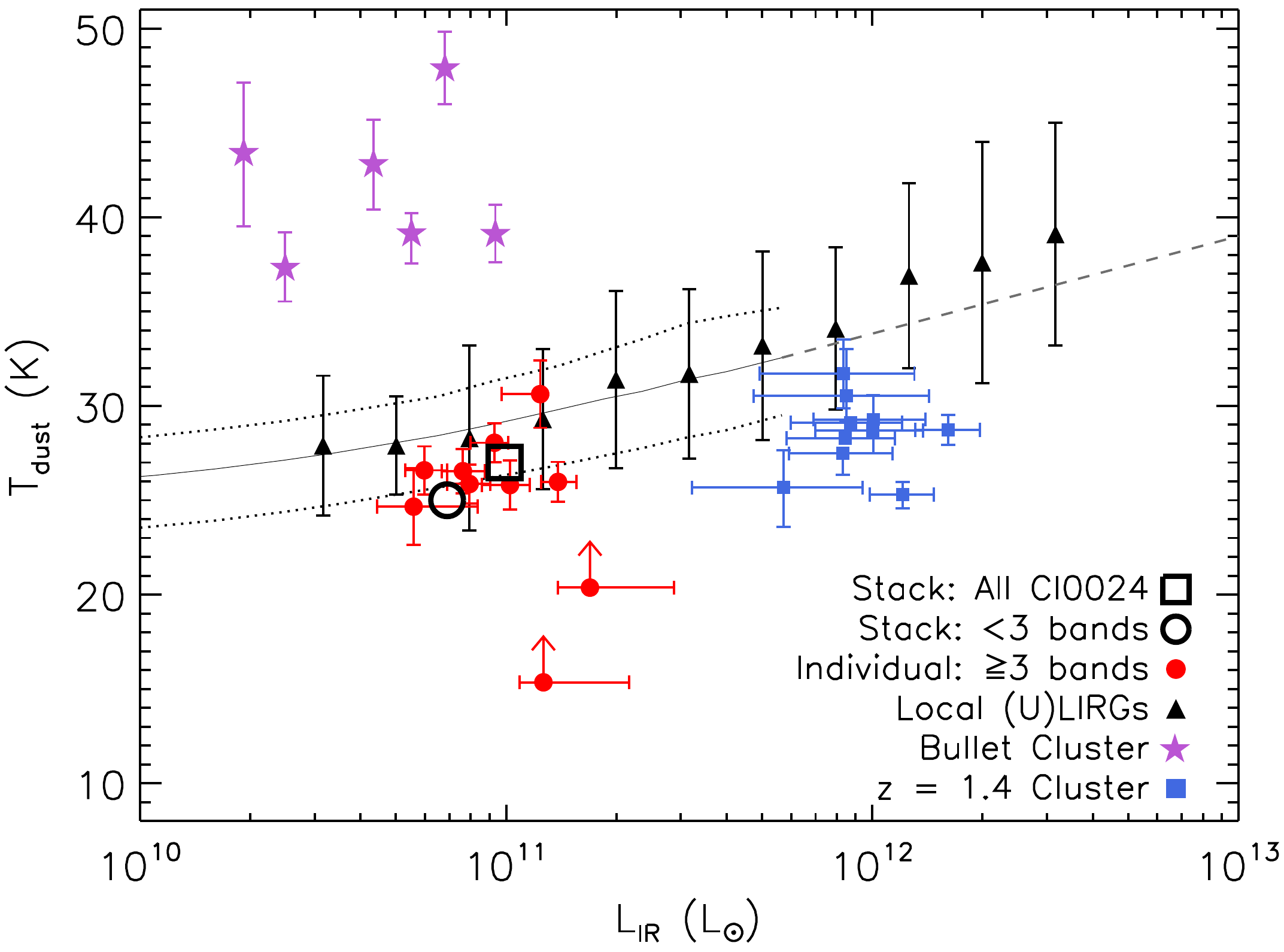}
  \caption{Characteristic dust temperature versus far-infrared
    luminosity for our cluster starburst sample, compared to the relation established for local
    starburst galaxies (black triangles show $z$\,<\,1 (U)LIRGs; \citealt{symeonidis2013}). Galaxies with detections in three or more infrared bands are shown in red points. We also show stacks of all 27 galaxies (open square) and those with fewer than three detections (open circle). We do not observe a significant offset from the local relation, in either direction. For comparison we plot the ``hot'' Bullet Cluster starbursts of \citet{rawle2012} which may have been stripped of their low density gas, and starbursts from a cluster at $z$\,=\,1.4 (which are consistent with high-redshift field galaxies; \citealt{ma2015}).}
    \label{fig:dust_temps}
\end{figure}

\subsection{Characteristic Dust Temperatures}
\label{dust}

The far-infrared observations also contain information regarding the
characteristic dust temperature.  Recenty, observations of local
clusters have revealed a population of apparently ``hot'' starbursts, with T$_{\rm d}\sim\,$\,50\,K and LIRG-like luminosities \citep{rawle2012}. These galaxies may represent a particular stage in the transition from
infalling spiral to cluster S0, where ram pressure has
stripped the low density (cold) gas disk but a central starburst is
yet to be quenched. In order to test whether the cluster starbursts in our sample are more or less evolved, we plot their far-infrared luminosity versus dust temperature in Fig.~\ref{fig:dust_temps}. For consistency with other comparision samples we employ a modified black-body curve of the form
$S_{\nu}\propto\nu^{\beta}$\,B$_{\nu}$T$_{\rm d}$ to derive the dust temperatures, with an emissivity index of
$\beta$\,=\,1.8.  We only include galaxies with detections in three
or more bands, which constitutes approximately a third of the sample. This
cut means that a reliable estimate of T$_{\rm d}$ can be derived.  

On the plot, we include the relationship for local LIRGs\,/\,ULIRGs (which is
derived from a sample of 60$\mu$m selected galaxies at $z\sim$\,0;
\citealt{chapman2003, chapin2009}) and SPIRE-selected $z$\,<\,1
(U)LIRGs \citep{symeonidis2013}, both of which follow a power-law
relation between L$_{\rm IR}\sim$\,10$^{10}$--10$^{13}$\,L$_{\odot}$. We also plot similar measurements of high-$z$ submillimeter-selected cluster
starburst galaxies in XCS\,J2215.9$-$1738 ($z$\,=\,1.4), which
show systematically colder dust temperatures at fixed luminosity
compared to the local relation \citep{ma2015}. This is likely to be a result of the more extended gas resevoirs compared to $z$\,=\,0 galaxies of the same luminosity.

For our sample of cluster starbursts we derive a median dust
temperature of $T_{\rm d}$\,=\,26\,$\pm$\,1\,K and far-infrared
luminosity $L_{\rm
  IR}$\,=\,(9.1\,$\pm$\,0.9)\,$\times$\,10$^{10}$\,$L_{\odot}$.  These
dust temperatures are consistent with those measured for low-redshift
field galaxies of similar far-infrared luminosity, and indeed we do
not find any starburst galaxies with dust temperatures exceeding
31\,K. However, we note that in this analysis we have only included the
galaxies with detections in three (or more) bands, and for two of these galaxies we place lower limits on T$_{\rm d}$ since we do not resolve the peak of the emission. To infer the average
dust temperature of the remaining sample we stack the PACS and SPIRE
images in four different ways: all 27 galaxies, galaxies detected in $\leq$\,1 band, in $\leq$\,2 bands and in $<$\,3 bands. The
characteristic dust temperatures of these subsets are all within 2\,K
of each other (25\,--\,27\,K), suggesting that the galaxies with well defined
blackbody fits are representative of the full of the sample. We plot the temperatures of the full sample stack and the $<$\,3 bands stack on Fig. \ref{fig:dust_temps} for
comparison. 

Finally, we note that although we only selected a small subsample for IFU follow-up, the 24$\mu$m MIPS parent sample comprises 60 galaxies in
total, approximately one third of which lie in the foreground group.
To search for differences between the group and cluster starbursts, we
stacked the PACS and SPIRE imaging of the 24$\mu$m-detected galaxies
in both subsets, but in each case the SPIRE colours and
characteristic dust profiles are consistent, with T$_{\rm
  d,cluster}$\,=\,23\,$\pm$\,2\,K and T$_{\rm
  d,group}$\,=\,23\,$\pm$\,3\,K respectively.

Thus, it appears that the characteristic dust temperatures of cluster starbursts are consistent with a
luminosity matched sample of field galaxies.  In this respect, our
sample appear to be less evolved than the ``hot'' starbursts
seen in some local clusters \citep{rawle2012}, perhaps indicating that they have been accreted more recently, and have not yet had their low density gas stripped by the ICM.

\begin{figure*}
    \centering
	\includegraphics[width=\textwidth, angle=0]{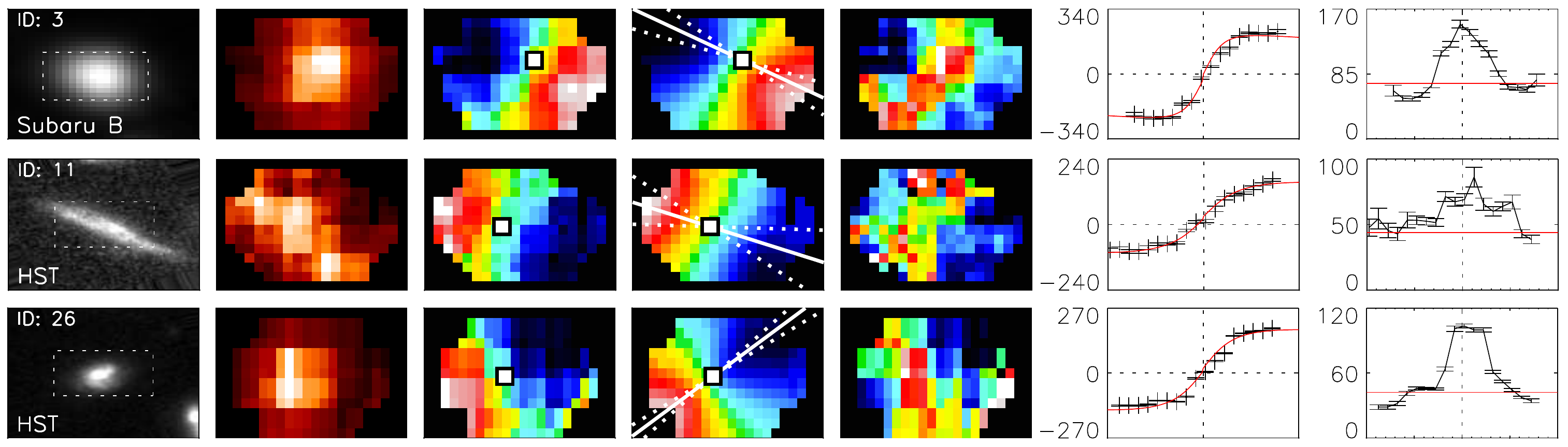}
    \caption{Example broad-band image, H$\alpha$ morphology and
      dynamics for three galaxies in our sample (data for the
      complete sample is shown in Fig.\ref{fig:dynamics_summary1}). From left to right:
      Subaru or \emph{HST} broad-band image, H$\alpha$ intensity map,
      H$\alpha$ velocity field, best-fit model velocity field,
      dispersion map, and the rotation curve and line-of-sight
      velocity dispersion profile extracted along the primary
      axis. The dashed outline on the broad-band image illustrates the
      position and size of the IFU
      (2$''$\,$\times$\,3$''$).  We fit the two dimensional data using
      a simple inclined disk model.  The solid line placed onto the
      model velocity field shows the position angle, and the dashed
      lines reflect the 1-$\sigma$ uncertainty of this. The square
      points indicate the position of the dynamical centre, according
      to the best-fit model.  We fit the one dimensional rotation
      curve (in red) to find a characteristic rotation velocity. Axes are in units of km\,s$^{-1}$.
      The solid red line in the last plot shows the value of
      $\sigma_0$. We find $\sim$\,90\% of our starburst sample resemble un-disturbed rotating disks, with a characteristic ``spider'' pattern in the velocity field, and a line of sight velocity dispersion which peaks towards the dynamical centre.}
    \label{fig:example_dynamics}
\end{figure*}

\subsection{Molecular Gas Masses}
\label{gas}
To determine how the molecular gas content of these galaxies compares to similar mass field galaxies, and calculate their likely final stellar mass if all available gas is converted to stars, we use the 
$^{12}$CO\,J(1\,$\rightarrow$\,0) emission.  We overlay the CO spectra
on top of the H$\alpha$ in Fig.~\ref{fig:onedspectra}, and find that the velocity centroid and line widths are well matched. The two sets of spectra have an average velocity offset of 50\,$\pm$\,10\,kms$^{-1}$, and an average ratio between the H$\alpha$ and CO line widths of 0.9\,$\pm$\,0.2. This
suggests that the $^{12}$CO traces the same dynamics as the H$\alpha$ emission, and that the ionised gas and molecular gas components are similarly
distributed throughout the disk.

Fitting each of these $^{12}$CO spectra with a single Gaussian profile, we find line luminosities of $L^{\prime}_{\rm CO}$\,=\,(2.2\,--\,6.8)\,$\times$\,10$^{9}$\,K\,km\,s$^{-1}$\,{\rm pc}$^2$ for the five detected galaxies \citep{solomon2005}, and upper limits of $L^{\prime}_{\rm CO}$\,=\,(3.9\,--4.9)\,$\times$\,10$^{9}$\,K\,km\,s$^{-1}$\,pc$^2$ for the rest of the sample (see Table \ref{table:derived}). To convert to the molecular gas mass we use $M({\rm
  H_2+He})\,=\,\alpha\,L^{\prime}_{\rm CO}$. The far-infrared
luminosities of our targets are much larger than what is typical of
local star-forming galaxies such as the Milky Way \citep{gao2004},
however as discussed in \citet{geach2011} the choice of
$\alpha$\,=\,4.6 may be the most appropriate given the late-type
morphologies of the sample, and that the average
$L_{\rm  IR}/L^{\prime}_{\rm CO}$
ratio we measure ($L_{\rm
  IR}/L^{\prime}_{\rm CO}$\,=\,25\,$\pm$\,5) is comparable to local
spirals. Using this calibration we estimate molecular gas masses of $M_{\rm
  gas}$\,=\,(1.0\,--\,3.0)\,$\times$\,10$^{10}$\,M$_{\odot}$ which
equates to gas fractions of $f_{\rm gas}$\,=\,$M_{\rm
  gas}$/($M_{\rm gas}$\,+\,$M_{\star}$)\,=\,0.11\,--\,0.39. As discussed in \S\ref{pdbi} we also stacked all 11 CO spectra to better characterise the sample as a whole. Using this stack we find a median line luminosity $L^{\prime}_{\rm CO}$\,=\,(2.7\,$\pm$\,0.3)\,$\times$\,10$^{9}$\,K\,km\,s$^{-1}$\,{\rm pc}$^2$, which translates to a gas mass of $M_{\rm
  gas}$\,=\,(1.2\,$\pm$\,0.2)\,$\times$\,10$^{10}$\,M$_{\odot}$ (towards the lower end of our detections). We will use this as a representative gas mass for our sample in all of the analysis below.
  
Our CO sample have an average star formation rate of (13\,$\pm$\,1)\,M$_{\odot}$\,yr$^{-1}$ (compared to $\sim$\,10\,M$_{\odot}$\,yr$^{-1}$ for the wider sample) and so we estimate a gas depletion timescale of $\sim$1\,Gyr, which is approximately $\sim$\,0.25 of the cluster crossing time. This may explain the non-detections of the $^{12}$CO in the (projected) cluster core. We predict final stellar masses for our sample with an average of $\sim$\,4\,$\times$\,10$^{10}$\,M$_{\odot}$. From a sample of local cluster S0s (0.04\,<\,$z$\,<\,0.07; WINGs survey) \citet{vulcani2011} find a characteristic stellar mass of M$_*$\,=\,2.2\,$\times$\,10$^{11}$\,M$_{\odot}$. This implies that our cluster starbursts are destined to be $\leq$\,M$_*$ members of the S0 population. It may be that the most massive $\leq$\,M$_*$ S0s are already in place at this redshift, or that their progenitors are passive spiral galaxies which have already completed their starburst phase.

As shown in Fig. \ref{fig:sample_properties}, our $^{12}$CO
sample includes galaxies of range of cluster-centric radii, and which inhabit both the ``main'' cluster and sub-group component. While all three
galaxies observed in the subgroup are detected in $^{12}$CO, only two
of eight galaxies are detected in the larger structure. This could be an environmental trend. However, we also caution that the
galaxies which we detect also
tend to be more far-infrared luminous, and given the correlation
between $L_{\rm IR}$ and $L^{\prime}_{\rm CO}$, this may simply mean the $^{12}$CO is easier to detect. Of course we cannot rule out the possibility that the same sources are brighter in the far-infrared due to some evolutionary effect. Considering our limited sample size it is difficult to draw definitive conclusions with respect to the gas content as a function of cluster
radius, but this study motivates a more detailed future study with a larger sample.

To compare the gas masses of these starbursts to galaxies on the star-forming ``main-sequence'' at this redshift, we use the
scaling relations of \citet{genzel2015}. These relations predict that a
star-forming galaxy at $z$\,$\sim$\,0.4 with a similar stellar mass will have a gas-to-stellar ratio of log(M$_{\rm gas}$/M$_{\star}$)\,$\sim$\,$-$0.9. This implies an average gas mass of M$_{\rm gas}$\,$\sim$\,4\,--\,8\,$\times$\,10$^9$\,M$_{\odot}$.  The molecular gas masses of our sample therefore appear to be slightly higher (a factor of $\sim$\,2) than those of ``typical'' galaxies of similar stellar mass at this redshift. The far-infrared and CO luminosities of our sample are consistent with the $L_{\rm IR}$\,--\,$L^{\prime}_{\rm CO}$ relation fit to low- and high-redshift star forming galaxies (including LIRGs, ULIRGs and SMGs) in \citet{ivison2013}.

Finally, we compare the gas properties of our cluster starbursts to starbursts in the field population, using a sample of ULIRGs
(L$_{\rm IR}$\,>\,10$^{12}$\,L$_{\odot}$) at 0.2\,<\,$z$\,<\,1 taken from
\citet{combes2013}.  This comparison sample have star formation rates $\sim$\,10$\times$ higher than our cluster starbursts, but by
mass matching their sample to the median stellar mass of our galaxy sample, we derive an average gas mass of M$_{\rm
  gas}$\,=\,(1.2\,$\pm$\,0.5)\,$\times$\,10$^{10}$\,M$_{\odot}$. This is consistent with the molecular gas masses derived for our cluster starburst sample.
  
On the basis of our $^{12}$CO observations it appears that these cluster starburst galaxies are richer in molecular gas than typical star forming galaxies at a similar redshift. We derive molecular gas masses which are more closely matched to those of starbursts (ULIRGs) in the field with similar infra-red luminosities. This could suggest a scenario in which the gas of infalling galaxies is compressed upon their encounter with the ICM, converting available H{\sc i} into H$_2$ and triggering a burst of star formation. These galaxies may have been typical star forming galaxies upon accretion to the cluster, yet have their molecular gas fraction enhanced through this process. We will return to these results in the final section.

\section{Analysis \& Discussion: Ionised Gas Dynamics}
\label{dynamics}
\subsection{Disk Fitting}
\label{fitting}

Now that we have established the integrated properties of our sample, we next investigate the spatially resolved properties as measured from the H$\alpha$
emission. The galaxies in our sample appear to resemble the field population
in terms of their stellar masses and characteristic dust temperatures (although possibly with higher molecular gas masses), and so our aim in this section is to assess how evolved these galaxies appear in terms of their gas dynamics. We will search for evidence of the mechanisms which may eventually quench the star formation in these galaxies and transform their kinematics from regular, disk-like rotation to pressure-supported S0s.

In Fig.~\ref{fig:dynamics_summary1} we show the two-dimensional maps
of H$\alpha$ emission, velocity, and line of sight velocity dispersion for all 27 targets. In each instance there is a clear velocity gradient
with peak-to-peak differences ranging from $\Delta
v$\,=\,100\,--\,500\,km\,s$^{-1}$. We note two galaxies (ID~23 and ID~24) exhibit very high $\log$([N{\sc ii}]/H$\alpha$) ratios (i.e. $\gtrsim$\,0); indicating the presence of an AGN (e.g. \citealt{kewley2013}). Additionally, these sources show irregular velocity dispersion profiles, with spatially offset broad emission ($\gtrsim$\,150\,km\,s$^{-1}$; Fig. \ref{fig:dynamics_summary1}), which may indicate that outflows are effecting the gas kinematics in these sources (e.g. \citealt{harrison2016}). For these reasons, we omit these two galaxies from the dynamical analysis below, leaving a sample of 25.

Next we assign a dynamical classification to each of the galaxies in our sample(Fig \ref{fig:group_sfr}; Table \ref{table:dynamics}). We classify 22 of 25 starbursts ($\sim$\,90\,\%) as rotationally supported, since their velocity map is regular (with a characteristic ``spider'' pattern), their rotation curve smooth, and they possess a line of sight velocity dispersion map which peaks towards the dynamical centre. The optical morphologies of these galaxies also appear smooth (and in some cases disk-like), with no evidence of multiple centres, tidal tails or merger activity. We class the remaining three galaxies as irregular, since they show complex dynamics in their two dimensional maps, rotation curves or dispersion profiles. This fraction of disks is consistent with field
surveys at these redshifts, which have suggested that massive
galaxies are the most well ordered at all redshifts up to $z\sim$\,1.5
(e.g. \citealt{kassin2012,wisnioski2015,stott2016}; but see also \citealt{puech2007}). In Fig. \ref{fig:group_sfr} we demonstrate that the kinematics do not appear to be correlated with far-infrared luminosity or $K$ band magnitude.

To provide a quantitative measure of the dynamics, and derive their
basic parameters (such as disk inclination, $i$ and hence true rotational
speed), we fit each each velocity field using a
simple disk model with arctan rotation curve \citep[e.g.\ ][]{courteau1997}.  In total
we fit for six different parameters: the dynamical centre ($x_{\rm
  c}$, $y_{\rm c}$), asymptotic velocity $v_{c}$, velocity at centre
of rotation $v_{0}$, turnover radius $r_{t}$, position angle $\theta$,
and disk inclination $i$.  Although all velocity fields show small
discrepancies the majority appear to be well fit by this simple
model, with $<$\,$v_{\rm data}$\,$-$\,$v_{\rm
  model}$\,$>$\,=\,26\,$\pm$\,15\,kms$^{-1}$. Example dynamical maps are shown in
Fig.~\ref{fig:example_dynamics}, and similar images for the full
sample are given in appendix Fig.~\ref{fig:dynamics_summary1}. We note that the position angle and dynamical centre returned by the model fit clearly pass through the regions of minimum and maximum rotation velocity. We also compare our inclination values to those derived from the $HST$ and Subaru optical morphologies. Although there is some degree of scatter, we find our conclusions in the following sections would be unchanged had we used inclinations derived directly from the images instead. Where the inclination of a particular galaxy is more uncertain, this is reflected in the larger errors assigned to dynamical parameters in Table \ref{table:dynamics}.

Using the best-fit disk parameters, we extract one-dimensional rotation curves by collapsing the velocity field along the major kinematic axis. It is clear from these data that a number of the rotation curves asymptote and turn over beyond $\sim$\,2kpc. To fully characterise this shape we fit the rotation curves with a model that includes a dark matter component. The velocity field can therefore be expressed as $v^2\,=\,v^2_{\rm d}\,+\,v^2_{\rm h}\,+\,v^2_{\rm HI}$, where the subscripts denote the stellar disk, dark matter halo and H{\sc i} gas disk respectively. For the stellar disk, we assume the stars follow an exponential surface density \citep{freeman1970} which is characterised by a disk mass and radius, and for the dark matter we assume $v_{\rm h}^2(r)$\,=\,$G$\,M$_{\rm h}(<r)$\,/\,$r$, with a dark-matter density profile that is described by a core density and radius. Further details of the modelling used are discussed in Swinbank et al.\ 2016 (in prep). Although we do not attempt to infer the stellar/dark halo fractions (due to strong degeneracies), this parameterisation allows us to improve our measurement of total rotation speed at the disk radius. While in most cases we could extract this velocity directly from the raw data, a model is useful for when this is not possible. In the following analysis we use inclinations from the two-dimensional disk model and rotation velocities from this one-dimensional rotation curve fit.

As a representative rotation velocity we choose to define $v_{2.2}$, the velocity at 2.2 $\times$ the half-light radius, since this
typically samples the rotation curve in a region where it is no longer
rising steeply. Of our IFU sample, ten are covered by deep $HST$ imaging and
the remaining 17 by Subaru (see Fig. \ref{fig:dynamics_summary1}), and
we use these images to establish the continuum half light radius ($r_{1/2}$) by fitting a series of ellipses at the galaxy position angle. We deconvolve for the seeing, and use the model rotation curve to estimate the velocity at this radius. We give this rotation speed and the corresponding half-light radius in Table \ref{table:dynamics}.

\begin{figure*}
 \includegraphics[width=\textwidth, angle=0]{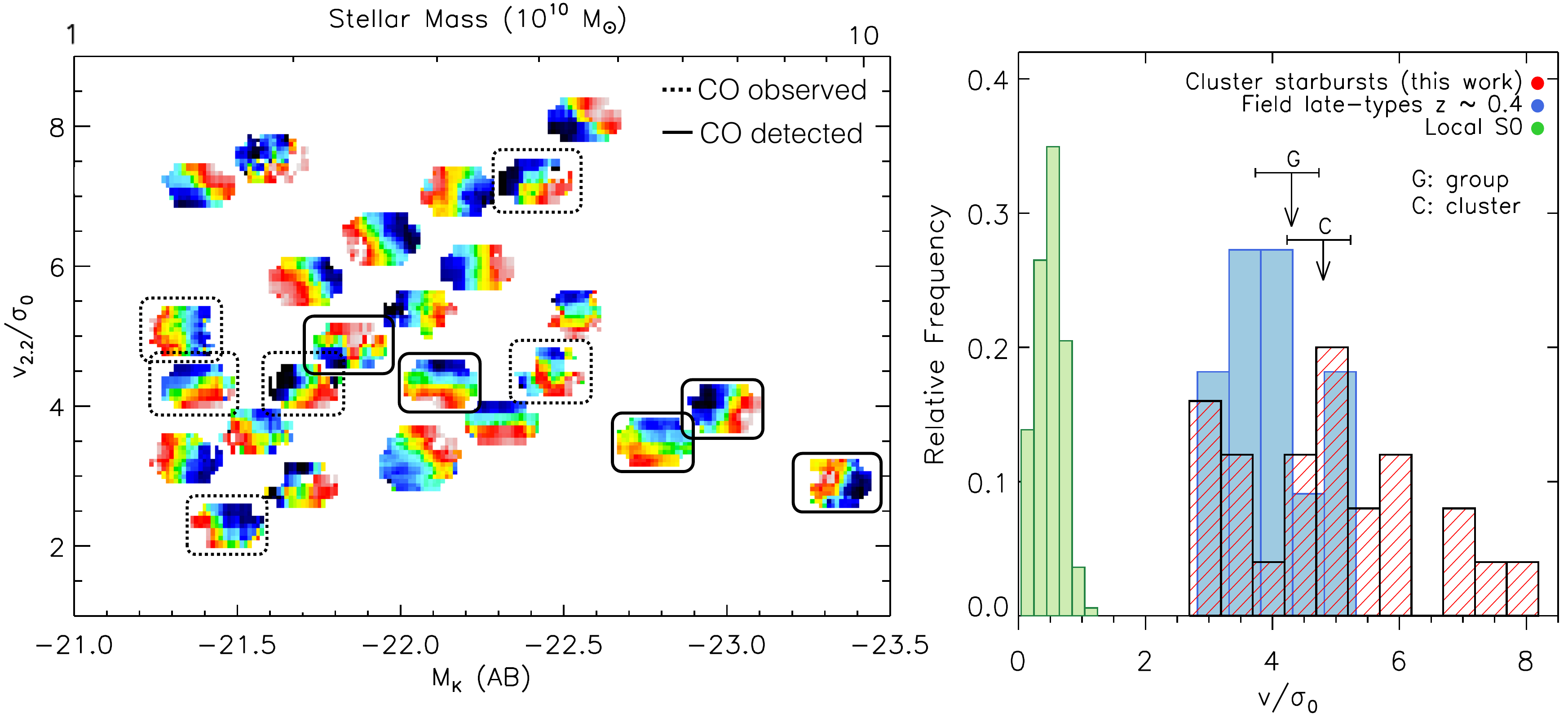}
  \caption{\emph{Left}: H$\alpha$ dynamics of the 25 galaxies as a function of rotational support ($v_{2.2}$/$\sigma_0$) and $K$ band magnitude. We choose to define the characteristic rotation velocity at 2.2 $\times$ the half-light radius, as measured from {\emph HST} and Subaru continuum images. We convert M$_{\rm K}$ values to stellar masses, using the average mass-to-light ratio derived from the {\sc hyperz} SED fitting (see \S3.1). We circle galaxies for which we have observed the $^{12}$CO transition, with solid lines corresponding to >\,5\,$\sigma$ detections, and dashed lines elsewhere. Some velocity fields have been shifted a small distance from their original position for clarity, however these values are provided in Tables \ref{table:derived} and \ref{table:dynamics}. The majority of galaxies resemble undisturbed, rotating disks, with no obvious trends between dynamics and galaxy mass or molecular gas content. \emph{Right}: The rotational support of our galaxies compared to local S0s \citep{emsellem2011} and field spirals of a similar redshift \citep{puech2007}. The fact that their $v$/$\sigma_0$ values are consistent with field spirals suggests that these galaxies have only recently been accreted. Their disks must be dynamically heated to decrease their $v$/$\sigma_0$ values to match S0s. The median $v$/$\sigma_0$ of the group does not appear to be significantly different to that of the ``main'' cluster ('G' arrow and 'C' arrow respectively).}
    \label{fig:poster_plot}
\end{figure*}

\subsection{Turbulence and Rotational Support}

To continue to address the question of which physical mechanism(s) drives
the star formation in cluster starbursts, we next measure the turbulence in the ISM. By comparing to galaxies in the field with
a mass and star formation rate matched sample, this will allow us to
quantify the effect of environment on the stability of the disk. To calculate
the intrinsic velocity dispersion, $\sigma_0$, we first correct for
beam-smearing effects using the raw two-dimensional velocity field.
For each pixel we consider all neighbouring pixels within the
PSF, and calculate the maximum velocity gradient across this element $\Delta v / \Delta R$.  This contribution to the dispersion is then removed and we
calculate the average velocity dispersion across the entire galaxy image
(inverse weighting by the noise at each pixel). The final corrected velocity
dispersions, $\sigma_0$, for each galaxy are given in Table
\ref{table:dynamics} and their values also shown relative to the
dispersion profiles in Fig.~\ref{fig:dynamics_summary1}. It can be seen that these $\sigma_0$ values are comparable to the velocity dispersion in the outer regions of the disk, which is less affected by beam smearing.

We find our cluster starbursts to have a median velocity dispersion of $\sigma_0$\,=\,50\,$\pm$\,15\,kms$^{-1}$. Dynamical studies between 0.2\,<\,$z$\,<\,1.0 report values in the range  $\sigma_0$\,=\,25\,--\,60\,kms$^{-1}$ (e.g. \citealt{kassin2012,wisnioski2015,stott2016}), however it is difficult to make direct comparisons since estimates are highly sensitive to the properties of the sample and the method used (see \citealt{stott2016} for further discussion). To derive $\sigma_0$ in a consistent way and compare to the turbulence of field galaxies at similar redshift, we exploit the sample of Swinbank et al. (2016, in prep.). This sample consists of $\sim$\,500 [O{\sc ii}] emitters serendipitously identified in a series of commissioning and science verification observations using MUSE. There were 16 ``extra-galactic'' fields observed, with the science targets largely ``blank'' fields or studies of high redshift ($z$\,>\,4) galaxies and quasars. We match the redshift, stellar mass and star formation rate of this star-forming sample to our cluster starbursts and find a median of $\sigma_0$\,=\,50\,$\pm$\,10\,kms$^{-1}$. This suggests that turbulence in the gas disks of our galaxies has not yet been enhanced by the various physical processes acting in this dense environment.

The ratio between inclination corrected rotation velocity and
intrinsic velocity dispersion, $v$/$\sigma_0$, is often used as a
measure of rotational support against thermal pressure.  We therefore
calculate this parameter for each galaxy in order to compare their
dynamics to the field population and cluster S0s, using the characteristic rotation velocities ($v_{2.2}$) calculated in \S\ref{fitting}. Across our sample we find inclination corrected values between $v_{2.2}$/$\sigma_0$\,=\,2.8\,--\,8.2, with a median of $v_{2.2}$/$\sigma_0$\,=\,5\,$\pm$\,2.  Given
that $v$/$\sigma_0$\,<\,1 is typically used as a cut-off for the
classification of ``dispersion dominated'' galaxies, this further
supports our conclusions that the majority of our sample are
undisturbed, rotating disks.

In Fig.~\ref{fig:poster_plot} we compare the rotation of the dusty starbursts to local S0s and spiral galaxies in low density environments at similar redshift. We plot the $v$/$\sigma_0$ values of 32 rotationally supported galaxies at 0.4\,$\le$\,$z$\,$\le$\,0.75 (also observed using FLAMES). \citet{puech2007} find a median $v$/$\sigma_0$\,=\,3.8\,$\pm$\,2, which is consistent with our sample. We also see that local S0s \citep{emsellem2011} typically have values of $v$/$\sigma_0$\,$\leq$\,1. The dynamics of infalling cluster galaxies must evolve significantly if a transformation is to take place between spirals and S0s. 

Finally, to search for trends in the dynamics as a function of
cluster-centric radius, we divide the sample into galaxies within the
(projected) cluster core, in the outskirts, and in the foreground sub-group.
However, we do not find any link between whether a galaxy resides in one of these regions and its dynamics (in terms of the observed dynamical
state, $\sigma_0$ or $v$/$\sigma_0$). We show the median $v$/$\sigma_0$ of these two subsamples in Fig. \ref{fig:poster_plot}. The kinematics also do not appear to be strongly correlated
with galaxy mass or star formation rate (see Fig. \ref{fig:group_sfr}). Although our results may
be complicated by projection effects, it
appears that the dynamics of the galaxies are not a strong function of
location within the cluster. This implies that any dynamical transformation must take
place over a longer period than the duration of the starburst ($\sim$\,1\,Gyr), or once the starburst has been quenched.

\vspace{-0.24cm}
\subsection{Specific Angular Momentum}

Although the ratio of $v$\,/\,$\sigma_{\rm 0}$ is a useful measure of the
dynamical state of a galaxy, a better quantification is the spin,
$\lambda_{\rm R}$, since this encodes how the rotational speed and
line-of-sight velocity dispersion vary with radius. Dynamical studies of
local galaxies have shown that early-type and late-type
galaxies tend to (broadly speaking) have different spin values \citep{querejeta2015,fogarty2015}.
\citet{emsellem2011} define $\lambda_{\rm R}$, which is essentially a proxy for specific angular momentum, as:

\begin{equation}
  \lambda_{\rm R} = \frac{\sum_{i=1}^{N}F_iR_i|V_i|}{\sum_{i=1}^{N}F_iR_i\sqrt{V_i^{2}+\sigma_i^{2}}}
\end{equation}
where $F_i$, $V_i$, $\sigma_i$, and $R_i$ are the flux, velocity,
velocity dispersion and radius of the $i$th pixel respectively. This
spin parameter was initially used in the classification of early-type galaxies, with \citet{cappellari2011} finding ellipticals and S0s could be split into two regimes -- fast rotators and slow rotators -- depending on their spin and ellipticity. However more recently the spin has been combined with the concentration of the stellar light profile, as a diagnostic tool to test how galaxies
might dynamically evolve from one type to another. Concentration, $c$, is defined as the ratio between the radii enclosing 90\% and 50\% of the Petrosian flux.

\begin{figure}
 \includegraphics[width=0.45\textwidth, angle=0]{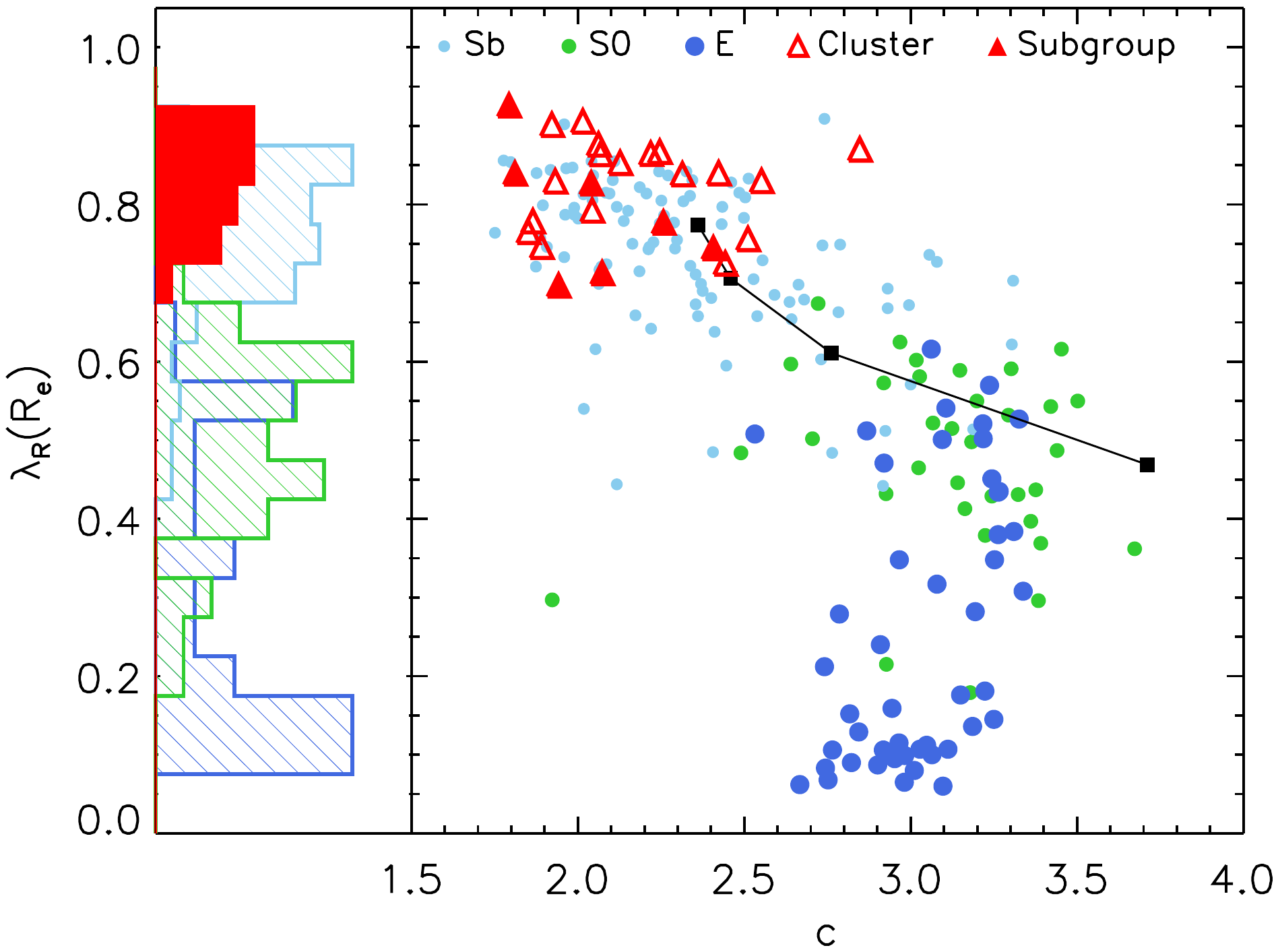}
  \caption{Galaxy spin, $\lambda_{\rm R}$ versus concentration, $c$
    for our starburst sample, as compared to galaxies of various morphological types from the CALIFA survey \citep{querejeta2015}. We also overlay predictions from the numerical models of \citet{bekki2011} who simulate the impact of tidal interactions on a typical spiral galaxy in a dense environment. Squares along the tracks represent 0, 2.8, 4.2 and 5.6\,Gyr since the beginning of this simulation. The high spin values of our galaxies suggest kinematics that are consistent with recently accreted field spirals. It seems that the mechanism which initiated the current burst of star formation has not had a dramatic effect on the disk dynamics. In order to make a transition between spirals and S0s these galaxies must lose angular momentum and increase their bulge to disk ratio (and therefore concentration).}
    \label{fig:lambda_r}
\end{figure}

In Fig.~\ref{fig:lambda_r} we compare the spin and concentration of
our sample to the properties of spirals and S0s from the CALIFA survey ($z$\,$\sim$\,0; \citealt{querejeta2015}). For consistency with the comparison samples we evaluate $\lambda_{\rm R}$ for all pixels within the half light radius,
deriving a median $\lambda_{\rm R}$\,=\,0.83\,$\pm$\,0.06 and
concentration of $c$\,=\,2.1\,$\pm$\,0.3 for the cluster galaxies. For this analysis we use \emph{HST} images (F814W filter) where available, and Subaru $z$-band images otherwise.
As Fig.~\ref{fig:lambda_r} shows, the dynamics and concentration of our galaxies are consistent with field spirals, implying that the dynamics of our sample are still relatively unaffected by the cluster environment. If these galaxies are to eventually
transition to S0s they clearly must undergo a process (or several)
which not only dynamically heats the disk and reduces rotational
support, but which also increases the bulge to disk ratio (and hence
increases the concentration by a factor $\sim$\,2).

Interactions with the dense ICM can effectively strip the gas
disks of spirals and lead to a rapid truncation of star formation. It was recently suggested that ram pressure stripping may even cause a temporary enhancement of star formation in central regions \citep{bekki2011,bekki2014}. However, the same numerical simulations predict little effect on the dynamics of these galaxies due to interaction with the ICM. Conversely, tidal interactions are very efficient at disrupting the disk. Repeat galaxy-galaxy encounters can act to increase the velocity dispersion and decrease the spin, potentially channelling gas inwards to fuel episodes of bursty star formation. Indeed, \citet{bekki2014} predict that a high spin, low concentration late-type galaxy in a high mass group may be transformed via tidal interactions into a low spin, high concentration S0, over a period of 2\,--\,4\,Gyr. We overlay this evolution in Fig. \ref{fig:lambda_r}. Although the timescales likely increase in a cluster environment, such a mechanism would simultaneously achieve the two key changes required for a spiral to S0 transition. In this scenario, the concentration is increased gradually via multiple bursts of star formation in the central regions, and we may expect to find starbursts with a range of $\lambda_{\rm R}$ and $c$ values. However, our cluster starburst sample are exclusively high-spin, low-concentration galaxies. We propose instead that the initial encounter with the ICM is responsible for triggering the starbursts observed in this cluster. Repeat galaxy-galaxy interactions may occur after the bursts have been quenched, to eventually achieve the $\sim$\,40\% decrease in angular momentum.

\subsection{The Beginning of the End?}
Overall, the characteristic temperatures, gas masses and dynamics of
these dusty starburst galaxies suggest that they have only recently been accreted from the field. The far-infrared star formation rates are on average twice that of a typical star-forming galaxy at this redshift, with our $^{12}$CO observations suggesting there is sufficient cold gas to sustain this activity for another $\sim$\,1\,Gyr. In fact we calculate molecular gas masses consistent with local ULIRGs. Dust temperatures in line with the local relation are further evidence that the gas disks have not yet been stripped by the intra-cluster medium.

The dynamics and morphologies of the cluster galaxies appear disk-like, with a large degree of rotational support ($v$/$\sigma_0$) and a high angular momentum ($\lambda_{\rm R}$). There is no evidence to suggest that the current bursts of star formation were initiated by mergers. It is possible that they were instead triggered by ram pressure, as an initial encounter with the ICM compresses the ISM of the infalling galaxies. This could also convert available H{\sc i} into molecular H$_2$, and may explain why these galaxies are so rich in star-forming gas. Such a proccess is consistent with the small excess of starbursts seen inside the virial radii of some clusters (e.g. \citealt{hogg2006}).

As we have seen in Fig. \ref{fig:lambda_r}, in order for the cluster starbursts to transition from spirals to S0s they must increase their concentration ($c$) and decrease their angular momentum ($\lambda_{\rm R}$). Multiple galaxy-galaxy interactions (harassment) seem a likely mechanism for the dynamical transformation, however the growth of the bulge does not appear to occur simultaneously. It is more likely that dynamical heating through tidal interactions occurs much later ($\geq$\,1\,Gyr), after the molecular gas has been depleted and/or stripped. The current burst looks set to increase the mass of these galaxies by $\sim$\,1\,$\times$\,10$^{10}$\,M$_{\odot}$, with final stellar masses consistent with sub-M$_*$ S0s. However, how this starburst episode may affect the bulge to disk ratio remains to be seen. To determine how much of the star formation is confined to central regions of the galaxy requires high resolution millimetre/submillimetre observations, for example using ALMA.

\section{Conclusions}
At low redshift the effects of environment on the morphology, gas content and star formation of galaxies is a well established phenomenon. Yet we know relatively little about the mechanisms which drive these trends, and how they depend on the properties of both the cluster, and the infalling galaxies on which they act. It is vital to develop a greater understanding of these issues if we are to explain the reversal of the morphology--density relation at high redshift. 

In this study we investigated the dynamics, star formation and gas properties of 27 cluster starbursts from Cl\,0024+17 ($z$\,$\sim$\,0.4), to learn more about the nature of these galaxies, and how dense environments at high redshift can promote star formation in some infalling galaxies. This problem is key to explaining the early and rapid formation of today's massive cluster galaxies. Our main findings are as follows:

\begin{itemize}
\item We use deblended \emph{Herschel} PACS\,/\,SPIRE maps to
  derive bolometric luminosities of $L_{\rm
    IR}$\,=\,(0.47\,--\,2.47)\,$\times$\,10$^{11}$ L$_{\odot}$ and
  star formation rates SFR$_{\rm
    IR}$\,=\,5\,--\,26\,M$_{\odot}$yr$^{-1}$.  The enhanced activity
  of these galaxies places them above the star forming
  ``main-sequence'' for this redshift and stellar mass (an average of
  $\sim$\,3\,$\times$\,10$^{10}$\,M$_{\odot}$).

\item From the far-infrared photometry, we derive characteristic dust
  temperatures of $T_{\rm d}$\,=\,26\,$\pm$\,1\,K, consistent with
  the local $L_{\rm IR}$-$T_{\rm d}$ relation for field galaxies.
  We do not find any evidence that the cold gas/dust has been
  stripped by the interaction of the galaxy ISM with the intra-cluster
  medium. If such an interaction occurs, then it must act on a much
  longer timescale, or occur after the initial burst.

\item We search for the $^{12}$CO\,J(1$\rightarrow$0) emission from
  11 galaxies in our sample. Of these targets, eight are within Cl\,0024+17 and three within the foreground
  sub-group. Only two of eight galaxies are detected in the larger
  structure, while all three galaxies in the
  foreground group are detected at >\,5$\sigma$. We find the
  average $^{12}$CO-derived gas mass of the sample (stacking detections and non-detections) to be $\sim$\,1\,$\times$\,10$^{10}$\,M$_{\odot}$. A median star formation rate of $\sim$\,13\,M$_{\odot}$\,yr$^{-1}$ suggests gas depletion timescales of $\sim$\,1\,Gyr, which is $\sim$\,0.25 of the cluster crossing time. Our galaxies appear to be richer in star forming gas than typical field galaxies at the same redshift.

\item We use FLAMES multi-IFU data to study the ionised gas dynamics
  of these cluster starbursts, as traced by H$\alpha$ emission, and after excluding AGN we find the majority of the sample ($\sim$\,90\%) have dynamics that appear to be consistent with undisturbed, uniformly rotating disks. To quantify the
  ratio between rotation and pressure support we calculate
  $v$/$\sigma_0$. The average for our sample is
  $v$/$\sigma_0$\,=\,5\,$\pm$\,2, in line with spirals of a similar redshift, further demonstrating that these galaxies have only recented accreted from the field.

\item We also measure the spin, $\lambda_{\rm R}$ (a proxy for specific angular momentum) and
  concentration, $c$ of the cluster starbursts. The relation
  between $\lambda_{\rm R}$ and $c$ provides a useful means to follow
  the morphological and dynamical evolution of these galaxies. We derive a
  median spin and concentration of $\lambda_{\rm R}$\,=\,0.83\,$\pm$\,0.06 and $c$\,=\,2.1\,$\pm$\,0.3, respectively. These values are consistent with typical field spirals. In order to evolve to S0s, these galaxies must double their concentration (through growth of the bulge) and decrease their angular momentum by a factor of $\sim$\,1.5.
  
  \item Although limited by small numbers, galaxies in the sub-group do not show evidence of an accelerated evolution. The only difference between the ``main'' cluster and sub-group components is (potentially) in the molecular gas properties, where all three sub-group members are detected in $^{12}$CO but only two of eight galaxies are detected in the larger structure.

\end{itemize}

It appears that these dusty, star-forming galaxies must have only recently been accreted to the cluster, with their dynamics, morphologies and molecular gas consistent with star forming field galaxies. They show no evidence of having yet been significantly disrupted by the dense environment. We conclude that for these $z$\,$\sim$\,0.4 cluster galaxies to make the transition to S0s they must undergo a dynamical heating of the disk, and an
increase in concentration. While ICM-related processes such as ram pressure stripping will truncate the gas disk, the full transformation from spiral to S0 is unlikely to be achieved by this process alone. Of the various available mechanisms, galaxy-galaxy encounters within the cluster seems most probable. Since $\sim$\,90\% of our sample display disk-like dynamics, it seems that this must occur after the initial burst has been quenched, and without an associated starburst.

\section*{Acknowledgements}

We are grateful to the anonymous referee for useful comments. FLAMES IFU data employed in this analysis is from VLT programs 089.A-0983 and 092.A-0135, and is available through the ESO archive. HLJ, CMH, AMS, RGB, and IRS gratefully acknowledge support from the STFC (ST/K501979/1, ST/I001573/1 and ST/L00075X/1). IRS also acknowledges  support from  an  ERC  Advanced  Investigator programme {\sc dustygal} (321334) and  a  Royal  Society Wolfson  Merit  Award. JEG acknowledges the support of the Royal Society. 




\bibliographystyle{mnras}
\bibliography{0024_v4} 



\appendix
\section{additional material}

\begin{figure*}
 \includegraphics[width=0.89\textwidth, angle=0]{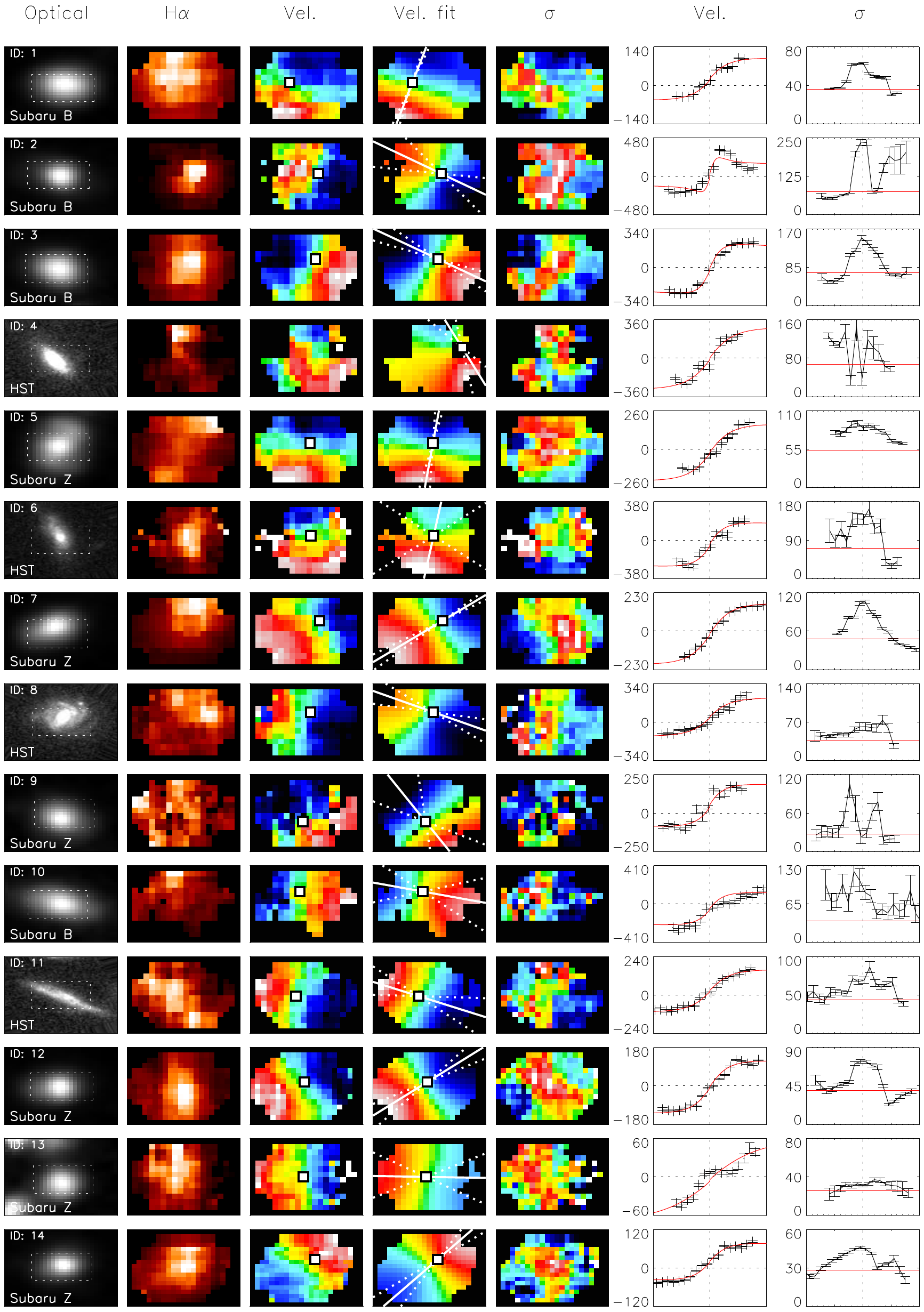}
  \caption{Broad-band image, H$\alpha$ morphology and dynamical maps for all 27 galaxies in our cluster starburst sample. Left to right we show: Subaru or \emph{HST} broad-band image (dashed lines illustrate the FLAMES field of view), H$\alpha$ intensity map, velocity field, best-fit model velocity field, dispersion map, and the rotation curve (red line is the model fit) and line of sight dispersion profile (red line is $\sigma_0$) extracted along the primary axis. Axes are in units of km\,s$^{-1}$. The solid line on the velocity field shows the position angle, with dashed lines illustrating the 1$\sigma$ uncertainty. Square points show the model dynamical centre.}
  \label{fig:dynamics_summary1}
\end{figure*}

\clearpage
\begin{figure*}
  \begin{center}{
  \includegraphics[width=0.89\textwidth, angle=0]{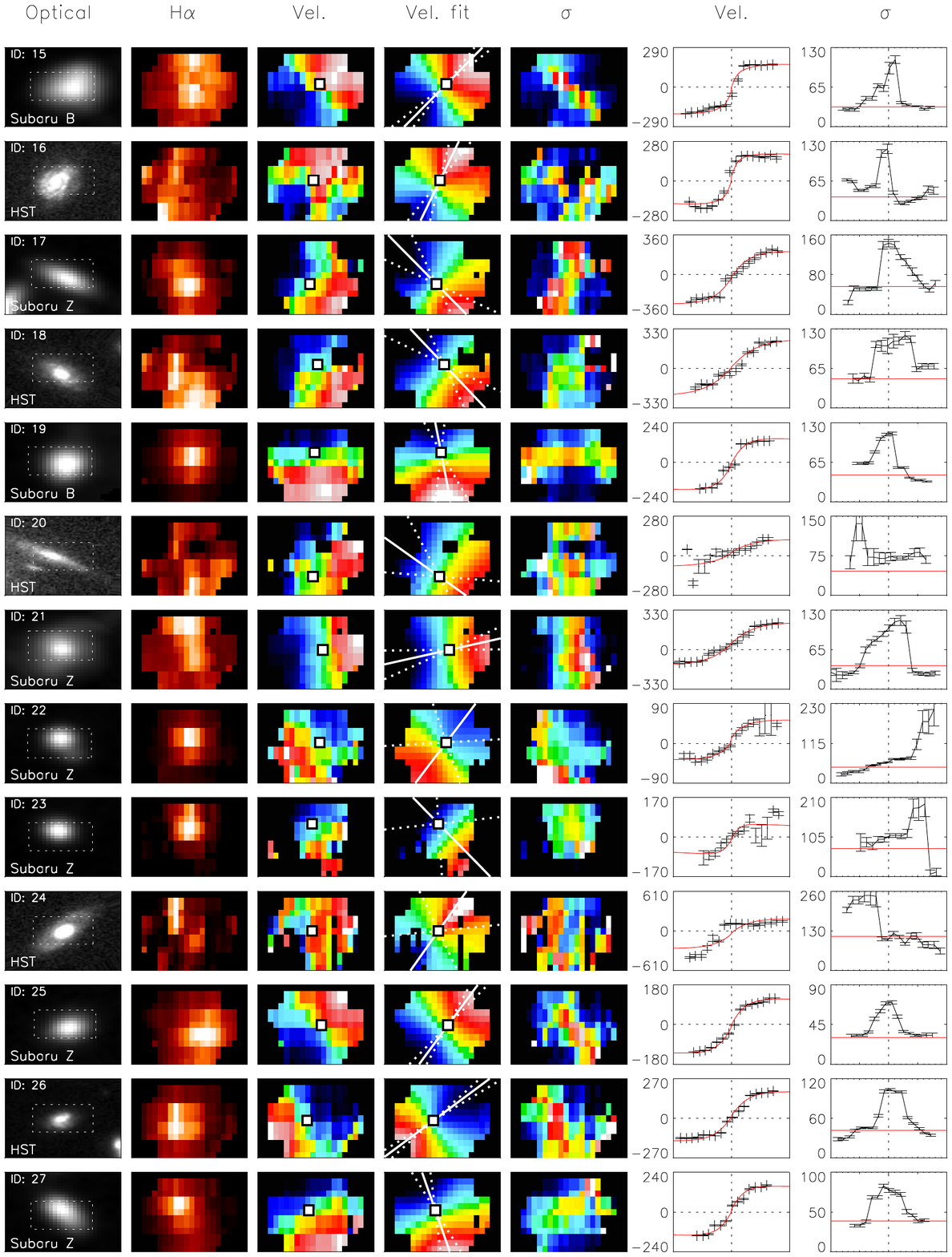}}\end{center}
  \contcaption{Broad-band image, H$\alpha$ morphology and dynamical maps for all 27 galaxies in our cluster starburst sample. Left to right we show: Subaru or \emph{HST} broad-band image (dashed lines illustrate the FLAMES field of view), H$\alpha$ intensity map, velocity field, best-fit model velocity field, dispersion map, and the rotation curve (red line is the model fit) and line of sight dispersion profile (red line is $\sigma_0$) extracted along the primary axis. Axes are in units of km\,s$^{-1}$. The solid line on the velocity field shows the position angle, with dashed lines illustrating the 1$\sigma$ uncertainty. Square points show the model dynamical centre.}
  \label{fig:dynamics_summary2}
\end{figure*}

\clearpage
\begin{figure*}
 \includegraphics[width=0.9\textwidth, angle=0]{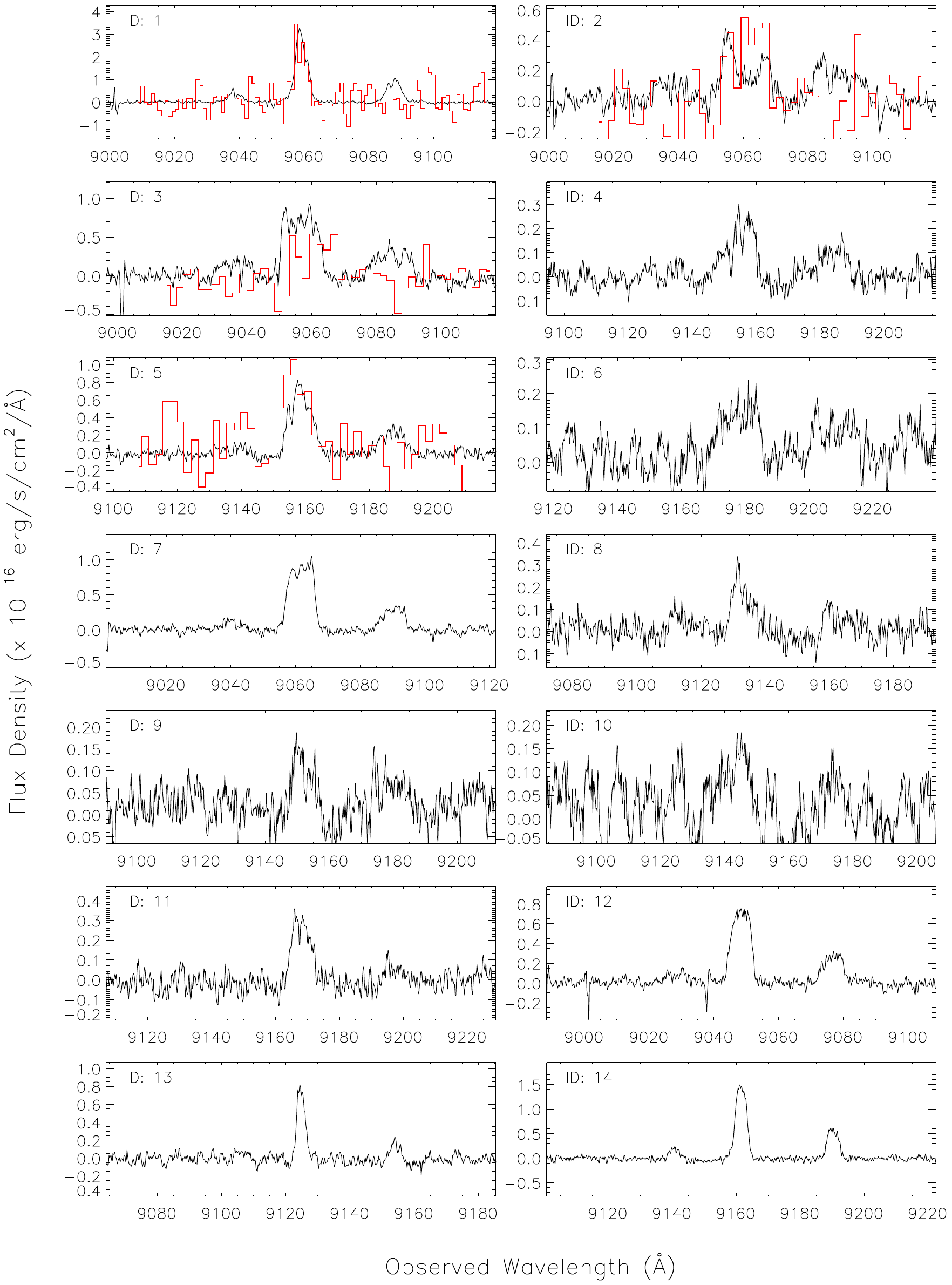}
  \caption{Integrated spectra around redshifted H$\alpha$ and [N{\sc ii}] (6548,6583) emission lines. Parameters such as redshifts and [N{\sc ii}]/H$\alpha$ ratios are listed in Table \ref{table:observed}. Many galaxies show split emission line profiles, indicative of high velocity rotation and possibly strong obscuration towards the dynamical centre. We suspect the dynamics of galaxy IDs 23 and 24 may be affected by an AGN. For IDs 1, 2, 3, 5 and 16 we overlay the CO spectra in red, and find that the velocity centroid and line width are well matched to the H$\alpha$ emission.}
  \label{fig:onedspectra}
\end{figure*}

\clearpage
\begin{figure*}
\begin{center}{
 \includegraphics[width=0.9\textwidth, angle=0]{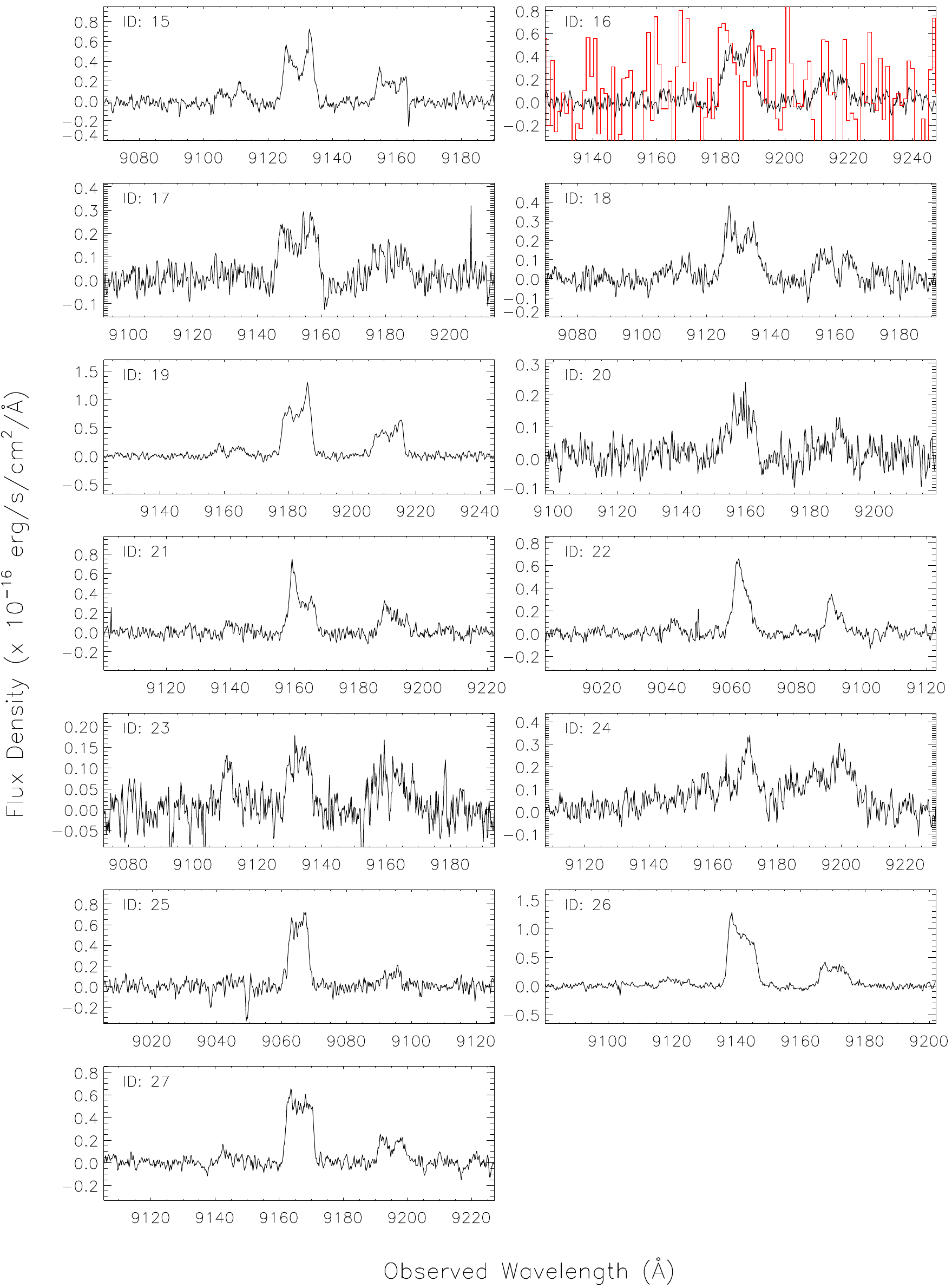}}
 \end{center}
  \contcaption{Integrated spectra around redshifted H$\alpha$ and [N{\sc ii}] (6548,6583) emission lines. Parameters such as redshifts and [N{\sc ii}]/H$\alpha$ ratios are listed in Table \ref{table:observed}. Many galaxies show split emission line profiles, indicative of high velocity rotation and possibly strong obscuration towards the dynamical centre. We suspect the dynamics of galaxy IDs 23 and 24 may be affected by an AGN. For IDs 1, 2, 3, 5 and 16 we overlay the CO spectra in red, and find that the velocity centroid and line width are well matched to the H$\alpha$ emission.}
  \label{fig:onedspectra2}
\end{figure*}

\clearpage
\begin{table*}
\begin{center}{
\caption{Observed Properties}
\begin{tabular*}{\textwidth}{l @{\extracolsep{\fill}} ccccccccccccc}
\hline

ID & RA & Dec & $z_{\rm H\alpha}$ & $S_{\rm 24}$ & $f_{\rm H\alpha}$ & log([N{\sc ii}]/H$\alpha$) &  $B$ & $R$ & Sub-group \Tstrut \\
& (J2000) & (J2000) & & ($\mu$Jy) & (10$^{-16}$\,erg\,s$^{-1}$) & & (AB) & (AB) & \Bstrut \\

\hline
\Tstrut

1  {\hskip 0.2in}  &    6.59050 &        17.3240 &       0.3804 &   950 $\pm$ 40     &        14.0   $\pm$   0.2 &     -0.48  $\pm$ 0.01  &    21.63  &    19.80     &        y \Bstrut \Bstrut \\
2    &    6.81270 &        17.2131 &       0.3804 &                 950 $\pm$ 50     &        4.2    $\pm$   0.2 &     -0.11  $\pm$ 0.03  &    22.53  &    19.98     &        y \Bstrut \Bstrut \\
3    &    6.71870 &        17.2332 &       0.3800 &                 830 $\pm$ 20     &        10.6   $\pm$   0.3 &     -0.37  $\pm$ 0.02  &    22.31  &    20.10     &        y \Bstrut \Bstrut \\
4    &    6.53730 &        17.2530 &       0.3950 &                 770 $\pm$ 30     &        2.2    $\pm$   0.1 &     -0.28  $\pm$ 0.04  &   23.88  &    21.32     &        n \Bstrut \Bstrut \\
5    &    6.76520 &        17.1912 &       0.3955 &                 706 $\pm$ 50     &        6.1    $\pm$   0.1 &     -0.44  $\pm$ 0.02  &   22.42  &    20.45     &        n \Bstrut \Bstrut  \\
6    &    6.60820 &        17.2976 &       0.3986 &                 440 $\pm$ 20     &        1.5    $\pm$   0.1 &     -0.24  $\pm$ 0.06  &    24.03  &    21.34     &        n \Bstrut \Bstrut \\
7    &    6.75580 &        17.3446 &       0.3808 &                 350 $\pm$ 30     &        8.5    $\pm$   0.2 &     -0.52  $\pm$ 0.01  &    22.16  &    20.48     &        y \Bstrut \Bstrut \\
8    &    6.66020 &        17.1254 &       0.3916 &                 201 $\pm$ 5      &        1.50   $\pm$   0.08 &    -0.38  $\pm$ 0.06  &    22.98  &    20.76     &        n \Bstrut \Bstrut \\
9    &    6.55990 &        17.2222 &       0.3944 &                 $<$ 150          &        0.83   $\pm$   0.07 &    -0.29  $\pm$ 0.07  &    23.36  &    21.15     &        n \Bstrut \Bstrut \\
10    &    6.66080 &        17.2539 &       0.3935 &                $<$ 150          &        0.72   $\pm$   0.08 &    -0.22  $\pm$ 0.07  &    23.28  &    20.95     &        n \Bstrut \Bstrut \\
11    &    6.64020 &        17.2055 &       0.3970 &                350 $\pm$ 50     &        2.30   $\pm$   0.08 &    -0.67  $\pm$ 0.06  &    23.57  &    21.71     &        n \Bstrut \Bstrut  \\
12    &    6.58330 &        17.0687 &       0.3788 &                670 $\pm$ 40     &        5.27   $\pm$   0.09 &    -0.42  $\pm$ 0.01  &    22.86  &    21.00     &        y \Bstrut \Bstrut \\
13    &    6.60210 &        17.1917 &       0.3904 &                290 $\pm$ 30     &        2.51   $\pm$   0.06 &    -0.58  $\pm$ 0.05  &    23.09  &    21.43     &        n \Bstrut \Bstrut \\
14    &    6.65550 &        17.0162 &       0.3960 &                600 $\pm$ 30     &        6.44   $\pm$   0.08 &    -0.39  $\pm$ 0.01  &    22.40  &    20.52     &        n \Bstrut \Bstrut  \\
15    &    6.67300 &        17.1838 &       0.3911 &                311 $\pm$ 5      &        4.7    $\pm$   0.2 &     -0.30  $\pm$ 0.04  &    22.26  &    20.10     &        n \Bstrut \Bstrut \\
16    &    6.63200 &        17.1545 &       0.3997 &                227 $\pm$ 4      &        5.1    $\pm$   0.2 &     -0.35  $\pm$ 0.03  &    22.86  &    20.80     &        n \Bstrut \Bstrut \\
17    &    6.66270 &        17.1672 &       0.3947 &                214 $\pm$ 4      &        2.5    $\pm$   0.1 &     -0.36  $\pm$ 0.05  &    23.91  &    21.42     &        n \Bstrut \Bstrut \\
18    &    6.63970 &        17.1566 &       0.3913 &                179 $\pm$ 4      &        3.1    $\pm$   0.1 &     -0.30  $\pm$ 0.03  &    23.01  &    20.80     &        n \Bstrut \Bstrut \\
19    &    6.69910 &        17.1619 &       0.3993 &                307 $\pm$ 7      &        8.9    $\pm$   0.2 &     -0.27  $\pm$ 0.02  &    22.13  &    20.17     &        n \Bstrut \Bstrut \\
20    &    6.52550 &        17.2054 &       0.3955 &                300 $\pm$ 50     &        1.33   $\pm$   0.08 &    -0.44  $\pm$ 0.06  &    23.92  &    21.71     &        n \Bstrut \Bstrut \\
21    &    6.74790 &        17.0626 &       0.3960 &                280 $\pm$ 30     &        3.8    $\pm$   0.1 &     -0.34  $\pm$ 0.03  &    22.53  &    20.51     &        n \Bstrut \Bstrut \\
22    &    6.63740 &        17.2581 &       0.3809 &                345 $\pm$ 30     &        3.05   $\pm$   0.06 &    -0.36  $\pm$ 0.02  &    23.35  &    21.29     &        y \Bstrut \Bstrut \\
23    &    6.72500 &        17.2343 &       0.3916 &                300 $\pm$ 30     &        1.11   $\pm$   0.07 &    -0.02  $\pm$ 0.04  &    23.79  &    21.03     &        n \Bstrut \Bstrut \\
24    &    6.58090 &        17.1174 &       0.3970 &                340 $\pm$ 30     &        3.1    $\pm$   0.2 &      0.04  $\pm$ 0.03  &   23.46  &    20.69     &        n \Bstrut \Bstrut \\
25    &    6.64270 &        17.1011 &       0.3813 &                260 $\pm$ 30     &        4.5    $\pm$   0.1 &     -0.69  $\pm$ 0.04  &    22.89  &    21.26     &        y \Bstrut \Bstrut \\
26    &    6.60710 &        17.2015 &       0.3929 &                340 $\pm$ 40     &        9.4    $\pm$   0.2 &     -0.46  $\pm$ 0.03  &    22.30  &    20.53     &        n \Bstrut \Bstrut  \\
27    &    6.65550 &        17.1768 &       0.3967 &                158 $\pm$ 5      &        5.2    $\pm$   0.1 &     -0.49  $\pm$ 0.03  &    23.11  &    21.21     &        n \Bstrut \Bstrut \\           

\hline

\end{tabular*}
\label{table:observed}
}\end{center}
{\sc notes:} The $B$- and $R$- band magnitudes were extracted at the same aperture, in order to derive the colours plotted in Fig. \ref{fig:sample_properties}. In the final column we specify whether each galaxy is a member of the ``main'' cluster (n) or foreground sub-group (y), as discussed in \S2.1. 
\end{table*}

\clearpage
\begin{table*}
 \begin{center}{
\caption{Derived Properties -- Galaxy Integrated}
\begin{tabular*}{\textwidth}{l @{\extracolsep{\fill}} cccccccc}
\hline

ID & L$_{\rm IR}$ & SFR$_{\rm IR}$ & T$_{\rm d}$ & M$_{\rm K}$ & $M_\star$ & $L^{\prime}_{\rm CO}$ & $M_{\rm gas}$ \Tstrut\Bstrut \\
   & (10$^{11}$\,L$_{\odot}$)  & (M$_{\odot}$\,yr$^{-1}$)  & (K) & (AB) & (10$^{10}$\,M$_{\odot}$) &  (10$^{9}$\,K\,km\,s$^{-1}$\,pc$^{2}$) & (10$^{10}$\,M$_{\odot}$) \Bstrut \\

\hline 
\Tstrut           

1  {\hskip 0.2in}  & 2.5 $^{+ 2  }_{- 3 }$        &  26  $^{+ 3}_{-13}$    & 29 $\pm$ 11  &   -22.8    & 5.6 $\pm$ 2.5  &    6.8 $\pm$ 0.6 &   3.1 $\pm$ 0.3   \Bstrut \Bstrut    \\
2                  & 0.5 $^{+ 0.9 }_{- 0.3 }$ *    &  11  $^{+ 2}_{-2}$      & $>$\,20    &   -23.4     & 9.6 $\pm$ 2.9 &   2.6 $\pm$ 0.3 &   1.2 $\pm$ 0.3    \Bstrut \Bstrut \\
3                  & 1.4 $^{+ 1.5 }_{- 1.2 }$     &  14 $^{+ 2}_{-1}$       & 25 $\pm$ 12 &   -23.0    & 6.8 $\pm$ 2.0  &  6.4 $\pm$ 0.5 &   2.9 $\pm$ 0.2    \Bstrut \Bstrut  \\
4                  & 1.3 $^{+ 2 }_{- 1 }$         &  13 $^{+ 9}_{-2}$       & $>$\,18     &   -22.5    & 4.1 $\pm$ 1.2  &  $<$\,3.9 &        $<$\,1.8      \Bstrut \Bstrut   \\
5                  & 0.6 $^{+ 0.8 }_{- 0.4 }$     &  6  $^{+ 3 }_{-1}$      & $>$\,23     &   -22.2    & 3.0 $\pm$ 1.4  &  4.3 $\pm$ 0.6 &   2.0 $\pm$ 0.3   \Bstrut  \Bstrut       \\
6                  & 0.5 $^{+ 0.9 }_{- 0.3 }$ *   &  10  $^{+ 2}_{-2}$      & 31 $\pm$ 22 &  -22.4      & 3.7 $\pm$ 1.1 &   ... &               ...       \Bstrut  \Bstrut  \\
7                  & 0.9 $^{+ 1.0 }_{- 0.8 }$     &  10  $^{+ 1}_{-2}$      & $>$\,11     &   -21.8    & 2.1 $\pm$ 0.8   & ... &               ...        \Bstrut  \Bstrut   \\
8                  & 0.8 $^{+ 0.9 }_{- 0.7 }$     &  8 $^{+ 1}_{-1}$        & 24 $\pm$ 2  &   -22.2    & 3.2 $\pm$ 1.0   & ... &               ...     \Bstrut  \Bstrut  \\
9                  & 0.8 $^{+ 0.9 }_{- 0.7 }$     &  8 $^{+ 1}_{-1}$        & $>$\,11     &   -21.7    & 1.9 $\pm$ 0.6   & ... &               ...  \Bstrut \Bstrut   \\
10                 & 0.6 $^{+ 0.7 }_{- 0.5 }$     &  6 $^{+ 1}_{-1}$        & $>$\,11     &   -22.1     & 2.9 $\pm$ 0.9  &  ... &               ...   \Bstrut   \Bstrut      \\
11                 & 1.6 $^{+ 3 }_{- 1 }$        &  17  $^{+ 13}_{-3}$      & 25 $\pm$ 1  &   -21.3    & 1.4 $\pm$ 0.5  &  ... &               ...    \Bstrut \Bstrut   \\
12                 & 1.0 $^{+ 1 }_{- 0.9 }$      &  11 $^{+ 1}_{-2}$        & $>$\,15     &   -21.9    & 2.5 $\pm$ 0.9  &  ... &               ...   \Bstrut  \Bstrut  \\
13                 & 0.5 $^{+ 1 }_{- 0.4 }$ *     &  3 $^{+ 1}_{-1}$         & 24 $\pm$ 2  &   -20.9    & 1.0 $\pm$ 0.4  &  $<$\,4.3 &        $<$\,2.0   \Bstrut        \Bstrut \\
14                 & 0.5 $^{+ 1 }_{- 0.4 }$ *    &  17  $^{+ 3}_{-3}$        & $>$\,11    &   -22.1     & 2.8 $\pm$ 1.1  &    ... &               ...    \Bstrut \Bstrut   \\
15                 & 0.6 $^{+ 2 }_{- 0.4 }$ *    &  9  $^{+ 2}_{-2}$       & 28 $\pm$ 1   &   -22.6     & 4.6 $\pm$ 2.1  &  ... &               ...   \Bstrut \Bstrut   \\
16                 & 0.5 $^{+ 1 }_{- 0.4 }$ *    &  11  $^{+ 2}_{-2}$        & 26 $\pm$ 1 &   -22.0     & 2.6 $\pm$ 1.2  &  2.2 $\pm$ 0.7 &   1.0 $\pm$ 0.3       \Bstrut  \Bstrut   \\
17                 & 0.5 $^{+ 2 }_{- 0.1 }$ *    &  12 $^{+ 2}_{-2}$        & 26 $\pm$ 1  &   -22.0    & 2.5 $\pm$ 1.2  &  $<$\,4.4 &        $<$\,2.0    \Bstrut  \Bstrut  \\
18                 & 0.5 $^{+ 0.9 }_{- 0.3 }$ *  &  5  $^{+ 1}_{-1}$        & 27 $\pm$ 2  &   -22.1     & 2.8 $\pm$ 0.9  &  $<$\,4.1 &        $<$\,1.9      \Bstrut    \Bstrut \\
19                 & 1.2 $^{+ 1 }_{- 1 }$        &  13  $^{+ 1}_{-3}$      & 26 $\pm$ 2   &   -22.7     & 5.0 $\pm$ 1.8 &   ... &               ...   \Bstrut  \Bstrut   \\
20                 & 0.9 $^{+ 5 }_{- 1 }$        &  20  $^{+ 31}_{-9}$     & $>$\,11      &   -21.7    & 2.1 $\pm$ 0.9  &  ... &               ...   \Bstrut \Bstrut   \\
21                 & 0.6 $^{+ 0.7 }_{- 0.5 }$    &  6 $^{+  1}_{-1}$        & $>$\,11     &   -22.3     & 3.4 $\pm$ 1.7 &  ... &               ...     \Bstrut \Bstrut    \\
22                 & 1.0 $^{+ 4 }_{- 0.9 }$      &  11 $^{+ 35}_{-1}$       & $>$\,11     &   -21.5    & 1.7 $\pm$ 0.8  &  ... &               ...  \Bstrut   \Bstrut   \\
23                 & 1.3 $^{+ 2 }_{- 1 }$        &  14  $^{+  7}_{-4}$      & $>$\,11     &   -21.7    & 2.0 $\pm$ 0.6  &    ... &               ...   \Bstrut  \Bstrut  \\
24                 & 1.3 $^{+ 2 }_{- 0.6 }$      &  13 $^{+  7}_{-6}$       & $>$\,11     &   -22.6    & 4.5 $\pm$ 1.4  &  ... &               ...   \Bstrut \Bstrut   \\
25                 & 0.5 $^{+ 0.5 }_{- 0.5 }$ *  &  6  $^{+ 2}_{-2}$       & $>$\,11      &   -21.0   & 1.1 $\pm$ 0.4  &  ... &               ...   \Bstrut \Bstrut   \\
26                 & 0.6 $^{+ 0.6 }_{- 0.6 }$    &  6  $^{+  1}_{-2}$       & 30 $\pm$ 2  &   -22.1    & 2.8 $\pm$ 1.3  &    $<$\,4.9 &        $<$\,2.3   \Bstrut       \Bstrut  \\
27                 & 0.5 $^{+ 0.9 }_{- 0.4 }$ *  &  8  $^{+  2}_{-2}$       & 27 $\pm$ 2  &   -21.3    & 1.3 $\pm$ 0.6  &  $<$\,4.6 &        $<$\,2.1   \Bstrut  \Bstrut  \\                

\hline
\end{tabular*}
\label{table:derived}
   }\end{center}
{\sc notes:} L$_{\rm IR}$ is the bolometric luminosity of the best-fit template SED fit to \emph{Herschel} far-infrared photometry. Star formation rates with an adjacent $*$ were derived by applying the average SFR$_{\rm H\alpha}$/SFR$_{\rm IR}$ correction factor, since these galaxies have detections in only $\leq$\,1 bands. The characteristic dust temperatures, T$_{\rm d}$, come from a single-temperature blackbody fit, assuming an emissivity index of $\beta$\,=\,1.8. Stellar mass estimates were calculated using a constant mass to light ratio of $M_{\odot}$/$L_{\odot}^K$\,=\,0.35. $L^{\prime}_{\rm CO}$ and $M_{\rm gas}$ are derived from a Gaussian fit to IRAM / NOEMA CO(1-0) spectra (see \S\ref{gas}), for which we assume a CO to H$_2$ conversion factor of $\alpha$\,=\,4.6.
\end{table*}

\clearpage
\begin{table*}
 \begin{center}{
\caption{Derived Properties -- Ionised Gas Dynamics}
\begin{tabular*}{\textwidth}{l @{\extracolsep{\fill}} cccccc}
\hline

ID & $\sigma_0$ & $r_{1/2}$ & $v_{2.2}$ & $v_{2.2}$/$\sigma_0$ & $\lambda_{\rm R}$ & Class \Tstrut\Bstrut \\
   & (km\,s$^{-1}$) & (kpc) & (km\,s$^{-1}$) & & & \Bstrut \\

\hline 
\Tstrut           

1  {\hskip 0.2in} & 36  $\pm$ 5     &    6.0  $\pm$  0.4     &   120  $\pm$  10    & 3.4   $\pm$ 0.5 &  0.71 $\pm$ 0.02   &   R   \Bstrut  \Bstrut  \\ 
2                 & 76  $\pm$ 8     &    6.2  $\pm$  0.5     &   210  $\pm$  40   & 2.8 $\pm$ 0.7    &  0.74 $\pm$ 0.01   &   R  \Bstrut  \Bstrut  \\
3                 & 72  $\pm$ 5     &    5.5  $\pm$  0.4     &   270  $\pm$  20   & 3.7 $\pm$ 0.5    &  0.77 $\pm$ 0.01   &   R  \Bstrut  \Bstrut  \\
4                 & 67  $\pm$ 5     &    8.5  $\pm$  0.4     &   300  $\pm$  20   & 4.4 $\pm$ 0.6    &  0.82 $\pm$ 0.01   &   I  \Bstrut  \Bstrut  \\
5                 & 53  $\pm$ 5     &    8.5  $\pm$  0.5     &   230  $\pm$  10   & 4.3 $\pm$ 0.5    &  0.82 $\pm$ 0.01   &   R  \Bstrut  \Bstrut  \\
6                 & 70  $\pm$ 9     &    9.6  $\pm$  0.5     &   310  $\pm$  20   & 4 $\pm$ 2        &  0.83 $\pm$ 0.01   &   R  \Bstrut  \Bstrut  \\
7                 & 48  $\pm$ 5     &    6.0  $\pm$  0.5     &   280  $\pm$  30   & 5.7 $\pm$ 0.9    &  0.83 $\pm$ 0.02   &   R \Bstrut  \Bstrut  \\
8                 & 37  $\pm$ 5     &    5.8  $\pm$  0.5     &   260  $\pm$  50   & 7 $\pm$ 2        &  0.90 $\pm$ 0.02   &   R  \Bstrut  \Bstrut  \\
9                 & 28  $\pm$ 7     &    5.9  $\pm$  0.4     &   210  $\pm$  60   & 8 $\pm$ 2        &  0.85 $\pm$ 0.03   &   R  \Bstrut  \Bstrut  \\
10                & 37  $\pm$ 7     &    9.5  $\pm$  0.4     &   190  $\pm$  20   & 5.3   $\pm$ 0.8  &  0.87 $\pm$ 0.01   &   R      \Bstrut  \Bstrut  \\
11                & 44  $\pm$ 5     &    9.8  $\pm$  0.5     &   140  $\pm$  10   & 3.2 $\pm$ 0.5    &  0.83 $\pm$ 0.01   &   R  \Bstrut  \Bstrut  \\
12                & 40  $\pm$ 5     &    5.0  $\pm$  0.4     &   250  $\pm$  50   & 6  $\pm$ 2       &  0.82 $\pm$ 0.07   &   R    \Bstrut  \Bstrut  \\
13                & 26  $\pm$ 5     &    5.4  $\pm$  0.6     &   130  $\pm$  40   & 5  $\pm$ 2       &  0.72 $\pm$ 0.02   &   R      \Bstrut  \Bstrut  \\
14                & 28  $\pm$ 5     &    6.9  $\pm$  0.2     &   100   $\pm$  10   & 3.4 $\pm$ 0.6   &  0.74 $\pm$ 0.03   &   R      \Bstrut  \Bstrut  \\
15                & 33  $\pm$ 5     &    6.0  $\pm$  0.4     &   300  $\pm$  20   & 8 $\pm$ 1        &  0.90 $\pm$ 0.02   &   R      \Bstrut  \Bstrut  \\
16                & 40  $\pm$ 4     &    5.7  $\pm$  0.4     &   200  $\pm$  10   & 5.0 $\pm$ 0.7    &  0.77 $\pm$ 0.04   &   R  \Bstrut  \Bstrut  \\
17                & 57  $\pm$ 7     &    6.8  $\pm$  0.8     &   290  $\pm$  30   & 5.1 $\pm$ 0.8    &  0.86 $\pm$ 0.01   &   R  \Bstrut  \Bstrut  \\
18                & 48  $\pm$ 5     &    6.0  $\pm$  0.6     &   290  $\pm$  30   & 6 $\pm$ 1        &  0.86 $\pm$ 0.05   &   R  \Bstrut  \Bstrut  \\
19                & 44  $\pm$ 4     &    4.2  $\pm$  0.3     &   220  $\pm$  40   & 5 $\pm$ 1        &  0.79 $\pm$ 0.03   &   R  \Bstrut  \Bstrut  \\
20                & 47  $\pm$ 6     &    8.8  $\pm$  0.7     &   130  $\pm$  30   & 2.8 $\pm$ 0.7    &  0.86 $\pm$ 0.01   &   I     \Bstrut  \Bstrut  \\
21                & 40  $\pm$ 4     &    9.7  $\pm$  0.4     &   230  $\pm$  20   & 5.9 $\pm$ 0.8    &  0.76 $\pm$ 0.03   &   R  \Bstrut  \Bstrut  \\
22                & 46  $\pm$ 5     &    5.2  $\pm$  0.5     &   150  $\pm$  20   & 3 $\pm$ 3        &  0.69 $\pm$ 0.02   &   I  \Bstrut  \Bstrut  \\
23                & 75  $\pm$ 6     &    4.0  $\pm$  0.4     &   90   $\pm$  30   & 1.2 $\pm$ 0.4    &  0.47 $\pm$ 0.01   &   I  \Bstrut  \Bstrut  \\
24                & 112 $\pm$ 7     &    8.0  $\pm$  0.3     &   230  $\pm$  20   & 2.0 $\pm$ 0.3    &  0.47 $\pm$ 0.03   &   I  \Bstrut  \Bstrut  \\
25                & 30  $\pm$ 5     &    6.0  $\pm$  0.4     &   220  $\pm$  20   & 7   $\pm$ 1      &  0.92 $\pm$ 0.05   &   R    \Bstrut  \Bstrut  \\
26                & 42  $\pm$ 4     &    3.7  $\pm$  0.5     &   190  $\pm$  10    & 4.5 $\pm$ 0.5   &  0.75 $\pm$ 0.06   &   R  \Bstrut  \Bstrut  \\
27                & 38  $\pm$ 5     &    5.8  $\pm$  0.4     &   200  $\pm$  30   & 5.2 $\pm$ 0.8    &  0.86 $\pm$ 0.04   &   R  \Bstrut  \Bstrut  \\

\hline
\end{tabular*}
\label{table:dynamics}
   }\end{center}
{\sc notes:} To estimate the intrinsic velocity dispersion, $\sigma_{0}$, we correct the two-dimensional $\sigma$ map for beam smearing and instrumental broadening, then take a pixel-by-pixel mean which is inverse weighted by the error. The rotational support, $v$/$\sigma_0$, is calculated using the value of the rotation curve at 2.2\,$\times$\,$r_{1/2}$ ($v_{2.2}$ and $r_{2.2}$ respectively), corrected for the inclination of the disk. The spin parameter, $\lambda_{\rm R}$, is calculated using all pixels within the half light radius (see \S4.3). The final column describes the dynamical class we assign, taking into account the two-dimensional dynamical maps, optical morphology, rotation curve and line of sight velocity dispersion profile -- 'R' if the galaxy resembles an undisturbed, rotating disk, and 'I' if the galaxy shows some sign of irregular kinematics or merger activity. We note that IDs 23 and 24 likely have an AGN which is disturbing the gas dynamics.
\end{table*}


\label{lastpage}
\end{document}